\documentclass[journal]{IEEEtran}

\ifCLASSINFOpdf
  \usepackage[pdftex]{graphicx}
\else
\fi
\usepackage{amsmath}
\usepackage{amsfonts}

\usepackage{multirow}
\usepackage{hyperref}
\usepackage{color}

\ifCLASSOPTIONcompsoc
 \usepackage[caption=false,font=normalsize,labelfont=sf,textfont=sf]{subfig}
\else
 \usepackage[caption=false,font=footnotesize]{subfig}
\fi

\begin{document}
\title{NAUTILUS: a Versatile Voice Cloning System}

\author{Hieu-Thi Luong,~\IEEEmembership{Student Member,~IEEE,}
        Junichi Yamagishi,~\IEEEmembership{Senior Member,~IEEE}
\thanks{H.-T. Luong is with the National Institute of Informatics, and with Department of Informatics, SOKENDAI (The Graduate University for Advanced Studies), Tokyo 101-8340, Japan (e-mail: luonghieuthi@nii.ac.jp)}
\thanks{J. Yamagishi is with the National Institute of Informatics, with Department of Informatics, SOKENDAI (The Graduate University for Advanced Studies), Tokyo 101-8340, Japan, and also with the Centre of Speech Technology Research, University of Edinburgh, Edinburgh EH8 9AB, U.K. (e-mail: jyamagis@nii.ac.jp).}
}

\markboth{IEEE/ACM Transactions on Audio, Speech, and Language Processing,~Vol.~XX, No.~XX, XX~2020}%
{Luong \MakeLowercase{\textit{et al.}}: NAUTILUS -- a Versatile Voice Cloning System}
%



\maketitle

\begin{abstract}
We introduce a novel speech synthesis system, called NAUTILUS, that can generate speech with a target voice either from a text input or a reference utterance of an arbitrary source speaker.
By using a multi-speaker speech corpus to train all requisite encoders and decoders in the initial training stage, our system can clone unseen voices using untranscribed speech of target speakers on the basis of the backpropagation algorithm.
Moreover, depending on the data circumstance of the target speaker, the cloning strategy can be adjusted to take advantage of additional data and modify the behaviors of text-to-speech (TTS) and/or voice conversion (VC) systems to accommodate the situation.
We test the performance of the proposed framework by using deep convolution layers to model the encoders, decoders and WaveNet vocoder.
Evaluations show that it achieves comparable quality with state-of-the-art TTS and VC systems when cloning with just five minutes of untranscribed speech. Moreover, it is demonstrated that the proposed framework has the ability to switch between TTS and VC with high speaker consistency, which will be useful for many applications.
\end{abstract}


\IEEEpeerreviewmaketitle

\section{Introduction}

\IEEEPARstart{S}{peech} synthesis is the technology of generating speech from an input interface. In its narrow sense, speech synthesis is used to refer to text-to-speech (TTS) systems \cite{tokuda2000speech}, which play an essential role in a spoken dialog system as a way for machine-human communication. In its broader definition, speech synthesis can refer to all kinds of speech generation interfaces like voice conversion (VC) \cite{kain1998spectral}, video-to-speech \cite{cornu2015reconstructing,michelsanti2020vocoder}, and others \cite{kinoshita2015text,hou2018audio,krishna2020speech}.
Recent state-of-the-art speech synthesis systems can generate speech with natural sounding quality, some of which are indistinguishable from recorded speech \cite{wang2017tacotron,shen2017natural}.
Deep neural networks are used in various components of these speech synthesis systems.
Many use sequence-to-sequence (seq2seq) models to unfold a compact phoneme sequence into acoustic features in the case of TTS \cite{shen2017natural,li2019neural} or to handle the misalignment of acoustic sequences in the case of VC \cite{miyoshi2017voice,tanaka2019atts2s,zhang2019non}. 
A neural vocoder, which generates waveforms sample-by-sample \cite{van2016wavenet,prenger2019waveglow,wang2019neural}, is also a staple of many high-quality speech-generation recipes \cite{shen2017natural,liu2018wavenet}. 
Generally speaking, the performance of deep learning approaches are high when training on a large amount of data. For speech generation models, this means that we need many hours of speech from a target speaker to train a model. This limits the ability to scale the technology to many different voices.

Besides improving the naturalness, cloning new voices with a small amount of data is also an active research topic.
While there are many different approaches proposed to tackle this problem, they all share the same fundamental principle which is using an abundant corpus to compensate for the lack of data of a target speaker \cite{jia2018transfer}. For neural TTS, we can fine-tune all or part of a well-trained acoustic model using transcribed speech from a target speaker \cite{inoue2020semi}. For neural VC, we can pool the speech data of multiple source and target speakers and share knowledge learned from each \cite{tian2018average}.
In most of these cases, the data used for training or adaptation is either paired or labeled. However, as all acoustic characteristics of a speaker are fully contained within speech signals, we should hypothetically be able to clone voices by using untranscribed speech only, and this would greatly reduce the cost of building speech generation systems. Disentangling speaker characteristics from linguistic information and representing it as a speaker vector is hence a popular way for cloning voices \cite{arik2018neural}.
Another approach is to use labels auto-generated by speaker-independent automatic speech recognition (ASR) trained on large-scale multi-speaker corpora \cite{inoue2020semi}. Either way, the cloning method is usually formulated for a specific data scenario of a specific speech generation system (either TTS or VC), while a true data-efficient method should work on extremely limited data and also abundant data with or without labels.

From the perspective of voice cloning, TTS and VC can be regarded as similar systems that use different inputs for generating speech with a target voice. They share almost the same objective as well as many functional components, but they are normally treated as different systems and are modeled using vastly different frameworks.
In this work, we present a novel speech generation system, called NAUTILUS, which can clone voices with a small amount of untranscribed speech and be used for both TTS and VC. It is expected to have state-of-the-art (SOTA) quality and highly consistent speaker similarity when switching between modes\footnote{The basis of the voice cloning method for TTS was proposed in \cite{luong2019unified} and as a proof-of-concept it was also shown that the same principle is applicable to VC in \cite{luong2019bootstrapping}. This new work builds upon the methodology and presents a SOTA unified voice cloning system for TTS and VC.}. More importantly, this combination has the ability to clone unseen voices with a versatile strategy that could be adjusted to accommodate the data situation of the target speakers.
Our experiments show that the proposed system is able to capture unique and subtle speaker characteristics such as L2 accents.

This paper is structured as follows: Section \ref{sec:relate} reviews works on TTS and VC in the context of cloning voices. Section \ref{sec:system} explains the principles 
of our framework. Section \ref{sec:condition} gives details on the NAUTILUS system used in this paper. Section \ref{sec:scenarios} presents experiment scenarios and their evaluations. We conclude our findings in Section \ref{sec:conclusion}.

\section{Related work on voice cloning}
\label{sec:relate}

\subsection{Definition of voice cloning}
The term \textit{voice cloning} is used to refer to a specific speaker adaptation scenario for TTS with untranscribed speech in several works \cite{arik2018neural,zhang2019learning}.
However in pop culture, it is loosely used to describe technology that resembles VC.
In this paper, we use \textit{voice cloning} as an umbrella term that indicates any type of system that generates speech imitating the voice of a particular speaker.
The main difference between \textit{voice cloning} and \textit{speech synthesis} is that the former puts an emphasis on the identity of the target speaker \cite{lorenzo2018can}, while the latter sometimes disregards this aspect for naturalness \cite{gutkin2016tts}.
Given this definition, a voice cloning can be a TTS, a VC, or any type of speech synthesis system \cite{michelsanti2020vocoder,krishna2020speech}. The NAUTILUS system is designed to be expandable to other input interfaces. However, we focus on TTS and VC, which are two common speech synthesis tasks, in this work as they play an irreplaceable role in our voice cloning framework.

The performance of a voice cloning system is judged on many aspects. As a speech generation system, the naturalness and similarity to target speakers are important \cite{shen2017natural}. As a computer system, a small memory footprint \cite{jia2018transfer} and fast computing time \cite{arik2018neural,tachibana2018efficiently} are desirable for practical reasons.
However, the defining property of voice cloning compared with generic speech synthesis is its data efficiency as this determines its scalability \cite{chen2018sample}.
While data efficiency can be interpreted as using as little data as possible \cite{jia2018transfer}, a better voice cloning system should not only work in a situation with extremely limited amount of data but also be able to take advantage of abundant speech data \cite{chen2018sample} when they become available regardless of the availability of transcriptions \cite{luong2019unified}. 

\subsection{Training voice conversion system for a target speaker}

The conventional VC approach is text-dependent, i.e.,\ it expects training data to be parallel utterances of source and target speakers \cite{stylianou1995statistical,toda2007voice}.
As obtaining these utterances is expensive and labor-intensive work, a parallel VC system has to commonly be built with as little as five minutes of data from a speaker \cite{lorenzo2018voice}. This is inconvenient and it limits the quality of VC systems in general.
Many have worked on methodologies for building VC systems with non-parallel utterances \cite{kameoka2018stargan}.
With HMM models, we can formulate a transformation function to adapt pretrained models using non-parallel speech \cite{chen2003voice,toda2006eigenvoice}.
With recent deep representation learning approaches, the popular method for non-parallel VC is training a speaker-disentangled linguistic representation either implicitly or explicitly. For implicit cases, Hsu \MakeLowercase{\textit{et al.}}\ \cite{hsu2016voice} used variational auto-encoder (VAE), while Kameoka \MakeLowercase{\textit{et al.}}\ \cite{kameoka2018stargan} used generative adversarial network (GAN) to train a \textit{many-to-many} non-parallel VC system. These methods use multi-speaker data, conditional labels, and various regularizations to encourage a model to disentangle linguistic content from speaker characteristics via a self-supervised training process. For explicit cases, Sun \MakeLowercase{\textit{et al.}}\ \cite{sun2016phonetic} used phonetic posteriorgrams (PPG) obtained from an ASR model to train an \textit{any-to-one} non-parallel VC system. As the ASR model is speaker-independent, a PPG-based VC system can theoretically convert the speech of arbitrary source speakers into a target speaker.
As PPG is a stand-in for text, the adaptation techniques used for TTS such as adapting an average \cite{yamagishi2007average} or a multi-speaker \cite{luong2017adapting} acoustic model can also be applied for PPG-based VC systems \cite{tian2018average}.

Even though a typical VC system is only trained on speech data, recent studies have suggested that using transcriptions of training data \cite{zhang2019non,wang2020end} or jointly training TTS along with VC \cite{zhang2019joint} can further improve naturalness of the generated speech.
In our previous work \cite{luong2019bootstrapping}, we established a methodology to bootstrap VC from TTS by utilizing the pretrained linguistic latent space. This paper builds upon this method by introducing an auxiliary phoneme recognition module and many new techniques to improve overall performance.

\subsection{Adapting text-to-speech system to an unseen target}

A TTS system is typically trained on dozens of hours of transcribed speech \cite{shen2017natural,ping2019clarinet}. Due to the high requirement for quantity and quality, a professional voice actor is commonly commissioned to record such data in a controlled environment. This makes the conventional approach ill-fitted for the voice cloning task in which we do not have controls over target speaker, recording environment, or the amount of data.
To build a TTS system for speakers with a limited amount of labeled data, we can adapt a pretrained model. The initial model can be trained on the data of a single speaker \cite{huang2018linear} or data pooled from multiple speakers \cite{yamagishi2007average,fan2015multi}.
This simple fine-tuning produces a high-quality model when the data of target speakers is sufficient (e.g.,\ one hour) \cite{chen2018sample}. When the data is extremely limited (e.g.,\ one minute), we can restrict the tuning to certain components instead of the entire network to prevent overfitting \cite{fan2015multi,luong2018scaling,chen2018sample}. In summary, speaker adaptation transfers knowledge learned from abundant data of one or multiple speakers to reduce demand on a target.

The costly part of the voice cloning system is the data collecting process, especially the transcription of speech.
Theoretically speaking, as speaker characteristics are self-contained within an utterance we should be able to clone voices without using text.
One practical approach is obtaining automatically annotated transcriptions using a SOTA ASR system \cite{inoue2020semi}. However ASR-predicted transcriptions contain incorrect annotations, which affects the performance of the adaptation. Moreover, this approach assumes that a well-trained ASR is obtainable for the target language, which makes it impractical for low-resource languages \cite{gutkin2016tts} or performing cross-language adaptation \cite{chen2009state,zhang2019learning}.
Given the disentanglement ability of deep learning models, another approach is to train a speaker-adaptive model conditioned on a speaker representation extracted from speech \cite{arik2018neural,takaki2018unsupervised}.
The speaker representation can be an \textit{i-vector} \cite{wu2015study}, \textit{d-vector} \cite{doddipatla2017speaker,jia2018transfer}, or \textit{x-vector} \cite{cooper2019zero}, which are all byproducts of speaker recognition systems.
This approach has a computational advantage in that it does not involve an optimization loop \cite{arik2018neural}. However, the drawback is its limited scalability; in other words the speaker similarity seems to stop improving when more than a few seconds of speech is used \cite{chen2018sample}.
The basis of our backpropagation-based unsupervised adaptation method, with high scalability, was proposed in previous publications \cite{luong2018multimodal,luong2019unified}. 
This paper tests the same method on a more elaborate and integrated speech generation system to refine the quality and speaker similarity of the target speakers.

\subsection{TTS as speech-chain component}

Even though TTS and ASR, two essential modules of spoken dialog systems, are placed at the two ends of the human-machine communication interface and compliment each other, historically, they are built independently under different frameworks \cite{tokuda2000speech,gales2008application}.
Recent end-to-end speech models have reduced the technical difference between TTS and ASR systems and opened up the possibility of integrating them into a single ecosystem. Tjandra \MakeLowercase{\textit{et al.}}\ \cite{tjandra2017listening} developed the Speech Chain model which consists of a TTS and ASR that consume each other's output as their own inputs. Karita \MakeLowercase{\textit{et al.}}\ \cite{karita2019semi} factorized TTS and ASR into encoders and decoders and then jointly trained them all together by putting a constraint on the common latent space. 
The purpose of these unified systems is combining resources and enabling semi-supervised training. 

Similar to the situation with ASR, several studies have tried to combine VC with TTS \cite{zhang2019joint,zhang2019non} or bootstrapping VC from TTS \cite{luong2019bootstrapping,huang2019voice,wang2020end}.
However, their focus was on leveraging a TTS-like system for VC \cite{zhang2019non,huang2019voice} or vice versa \cite{zhang2019joint,polyak2019attention} in a data-abundant scenario (target speakers with a reasonable amount of transcribed speech), and they disregard the data efficiency aspect as well as the application synergy between the two systems.
Hypothetically speaking, given a perfect ASR system, there is no difference between TTS and VC systems. Specifically, the PPG-based VC system \cite{sun2016phonetic} is essentially a TTS model stacked on top of an ASR model.
Polyak \MakeLowercase{\textit{et al.}}\ \cite{polyak2019attention} trained a TTS with the target voice by combining any-to-one VC and robot-voice TTS systems \cite{polyak2019attention}.
In this paper, we focus not only on improving the performance of TTS and VC individually but also on developing a unified system which can perform both tasks with high consistency. Such systems would be useful for many practical application scenarios.

\section{Versatile voice cloning framework}
\label{sec:system}

Our proposed system is a multimodal neural network \cite{kinoshita2015text,hou2018audio}, that can be used as TTS \cite{luong2019unified} or VC \cite{luong2019bootstrapping}.
It is not just a combination of conventional TTS and VC systems \cite{zhang2019joint} but a carefully designed system that has the ability to clone unseen voices using either transcribed or untranscribed speech \cite{luong2019unified}. 
The core concept is to train a latent linguistic embedding (LLE) to use as a stand-in for text when transcription is difficult to obtain.
The architecture of our multimodal system resembles the model proposed by Karita \MakeLowercase{\textit{et al.}}\ \cite{karita2019semi}; however, they focus on the performance of ASR system instead of speaker adaptation. 
While the emphasis on linguistic latent features is similar to the PPG-based VC system proposed by Sun \MakeLowercase{\textit{et al.}}\ \cite{sun2016phonetic}, their phonetic representation extractor is trained independently with the VC model while our linguistic latent features are jointly trained with the speech generation model. 
Given the similarity in techniques, we will compare our system with the PPG-based VC system in the experiments.

\subsection{Training the text-speech multimodal neural network}

\begin{figure}[tb]
  \centering
  \includegraphics[width=0.95\columnwidth]{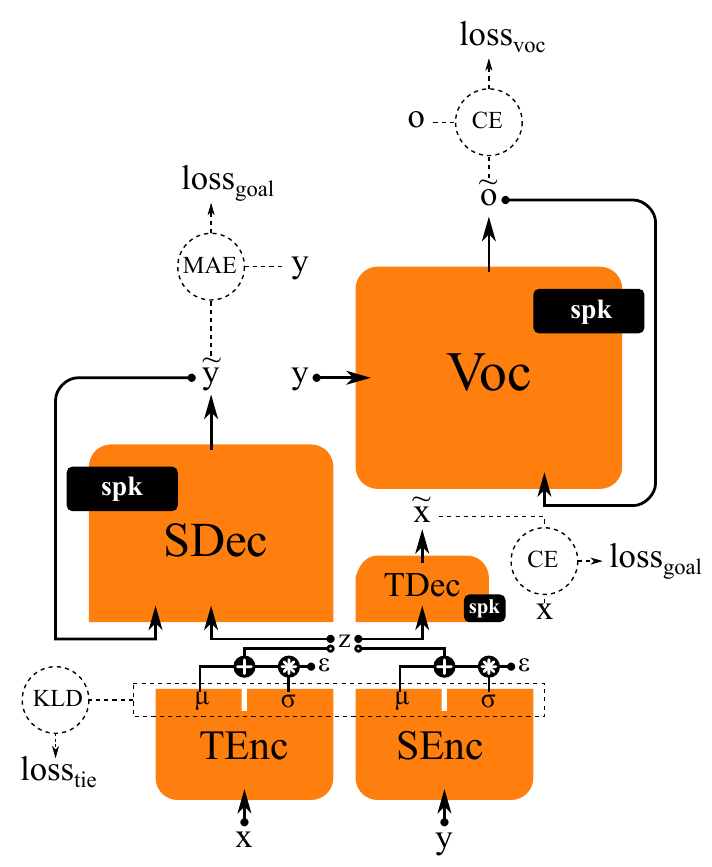}
\vspace{-3mm}
\caption{The proposed system comprises of a text encoder ($TEnc$), a speech encoder ($SEnc$), a text decoder ($TDec$), a speech decoder ($SDec$), and a neural vocoder ($Voc$). Where $\boldsymbol{x}$ is a text (phoneme) representation, $\boldsymbol{y}$ is a speech (acoustic) representation, $\boldsymbol{o}$ is a waveform representation, while $\boldsymbol{z}$ is a latent linguistic embedding. $\boldsymbol{\tilde{x}}$, $\boldsymbol{\tilde{y}}$, and $\boldsymbol{\tilde{o}}$ are approximations of the respective representations produced by the neural networks. $\textrm{loss}_{goal}$ is a placeholder for either $\textrm{loss}_{tts}$, $\textrm{loss}_{sts}$, $\textrm{loss}_{stt}$, or $\textrm{loss}_{ttt}$ depending on the encoder/decoder combination. Specific to the experiments in this paper, the speech generation tasks use the mean absolute error (MAE) while the speech recognition tasks use cross entropy (CE) as a cost function. CE is also used to for $\textrm{loss}_{voc}$ to train the neural vocoder, while the KL divergence (KLD) is used as the latent tying loss $\textrm{loss}_{tie}$. The black box with the word ``spk'' indicates the module contains speaker-dependent components. The encoders output the mean ($\mu$) and standard deviation ($\sigma$) of the latent features and then generate the features by using a random value ($\epsilon$) drawn from a standard normal distribution.}
\vspace{-3mm}
\label{fig:architecture}
\end{figure}

The main components of the framework are presented in Fig.\ \ref{fig:architecture}. The multimodal neural network is essential for our voice cloning methodology. While the neural vocoder is optional, we included it since it is necessary for generating high-quality speech in most recent setups \cite{shen2017natural,liu2018wavenet}.
The proposed system contains four modules, which are encoders and decoders of either text, $\boldsymbol{x}$, or speech, $\boldsymbol{y}$. In combinations of encoders and decoders, the modules can perform four transformations: text-to-speech (TTS), speech-to-speech (STS), speech-to-text (STT), and text-to-text (TTT).
These modules have synergies when trained together.
The speech encoder helps the TTS system adapt with untranscribed speech \cite{luong2019unified}, while the text encoder helps the VC system disentangle speaker from the linguistic \cite{luong2019bootstrapping}. 
The text decoder is the new addition in this paper. While Karita \MakeLowercase{\textit{et al.}}\ \cite{karita2019semi} use a similar setup for speech recognition, we focus on speech generation and the text decoder is only used as a regularizer.

\begin{figure*}[t]
  \centering
  \captionsetup[subfigure]{justification=centering}
  \subfloat[Step 1 - Adaptation\label{fig:step-heatup}]{\includegraphics[width=0.33\linewidth]{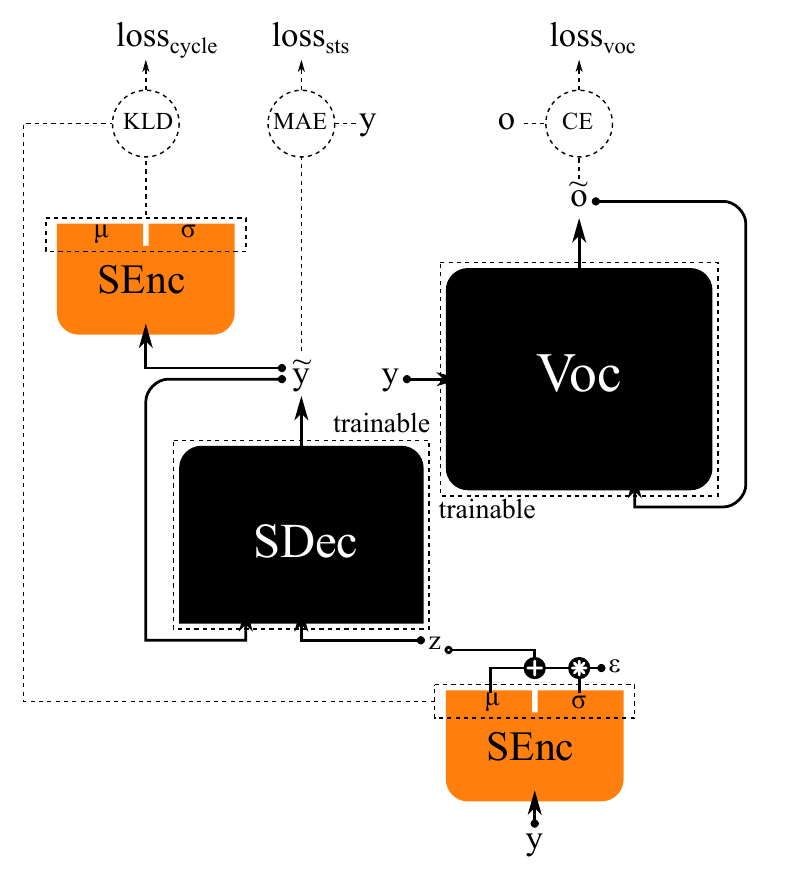}}
  \hfil
  \subfloat[Step 2 - Welding\label{fig:step-welding}]{\includegraphics[width=0.33\linewidth]{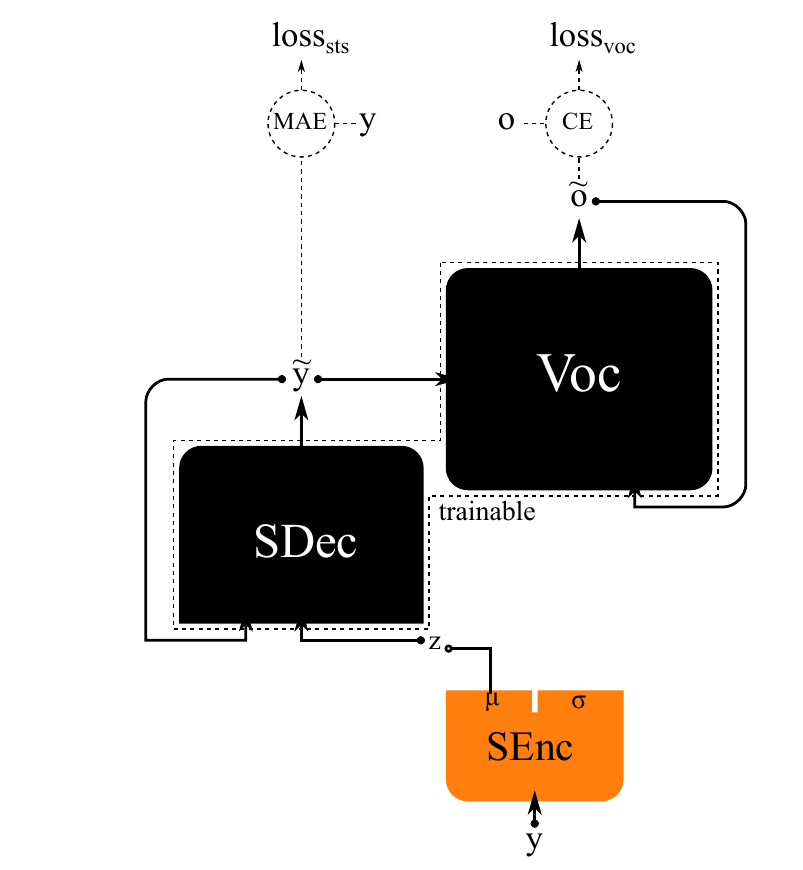}}
  \hfil
  \subfloat[Step 3 - Inference\label{fig:step-inference}]{\includegraphics[width=0.33\linewidth]{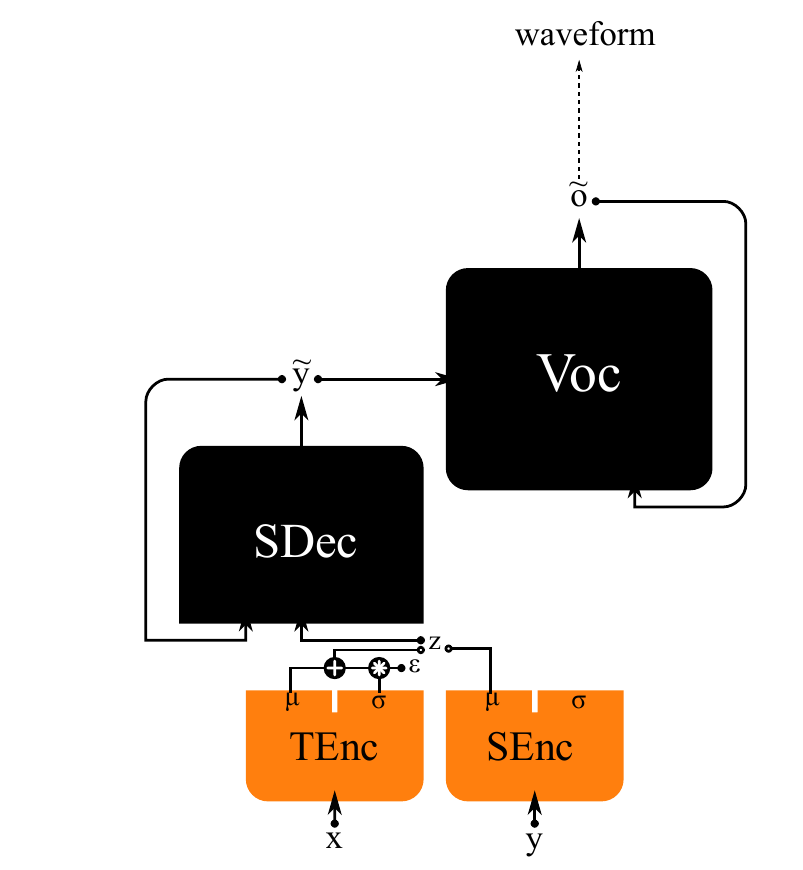}}
\caption{Cloning procedure with untranscribed speech of the target speaker. The black background indicates modules that were or will be adapted to target speaker's data, while the orange background indicates modules that were trained on a multi-speaker corpus in the training stage, and are, supposedly, universal. The trainable modules in each step are indicated by the dashed border with the word ``trainable'' near it. In the \textit{welding} and \textit{inference} step, the mean-value LLE tactic is applied to the speech encoder by assigning zero to the $\epsilon$ instead of sampling from a normal distribution.}
\vspace{-3mm}
\label{fig:stages}
\end{figure*}

Our methodology is designed around the training of a speaker-disentangled LLE, $\boldsymbol{z}$. The LLE in our setup plays the same role as the PPG proposed for VC \cite{hsu2016voice}. However, the LLE is jointly trained with the speech generation modules and contains linguistic information as a whole (instead of phoneme).
There are several ways to train the multimodal neural network. It can be trained stochastically \cite{li2016multi}, step-by-step \cite{huang2019voice}, or jointly \cite{luong2018multimodal,karita2019semi}.
We proposed two methods for the joint training in our previous work \cite{luong2018multimodal}: 1) \textit{joint-goal} where several losses calculated between an output inferred by each decoder and its ground-truth are combined, and 2) \textit{tied-layer}, where the distance or distortion between two latent spaces obtained from encoders are constrained to be identical. Using one or the other is enough \cite{luong2018multimodal,luong2019unified}, but as they are complementary, we could use them together:

\begin{equation}
\label{eq:losstrain}
\textrm{loss}_{train} = \textrm{loss}_{goals} +  \beta \; \textrm{loss}_{tie}
\end{equation}

Here, $\textrm{loss}_{goals}$ is a weighted combination of several types of $\textrm{loss}_{goal}$, which is a placeholder for the training losses created by combining different encoders and decoders of the multimodal networks. Specifically, given the text-speech multimodal system illustrated in Fig.\ \ref{fig:architecture}, we used the following equation as the joint-goal loss to train the initial model:
\begin{equation}
\label{eq:lossgoals}
\textrm{loss}_{goals} = \textrm{loss}_{tts} + \alpha_{sts} \; \textrm{loss}_{sts} + \alpha_{stt} \; \textrm{loss}_{stt} \; ,
\end{equation}
where $\textrm{loss}_{tts}$ is the TTS loss defined by the text encoder and speech decoder, it is also used as the anchor to adjust other hyperparameters; $\textrm{loss}_{sts}$ is the STS loss defined by the speech encoder and speech decoder, it is de-emphasized by the weighting parameter $\alpha_{sts}$;
and $\textrm{loss}_{stt}$ is the STT loss defined by the speech encoder and text decoder, even though speech-to-text is not a target task, $\textrm{loss}_{stt}$ is included to encourage the latent space to focus more on phonemes (but not entirely). Some other works have shown that an auxiliary phoneme classifier helps in boosting the quality of speech generation systems in general \cite{zhang2019non}. The TTT loss defined by the text encoder and text decoder, $\textrm{loss}_{ttt}$, is not included as we do not think that it helps.

In each training step, we calculate each term of the loss$_{train}$ using a transcribed speech sample and then optimize all parameters in a supervised manner. Karita \MakeLowercase{\textit{et al.}}\ \cite{karita2019semi} used a similar loss to jointly train their system but with one important difference, 
two separated speech samples, one with its transcription and another without, are used to calculate a single training loss. Specifically, $\textrm{loss}_{tts}$,  $\textrm{loss}_{stt}$, and $\textrm{loss}_{tie}$ are calculated using the transcribed sample, while $\textrm{loss}_{sts}$ and $\textrm{loss}_{ttt}$ are calculated on the untranscribed sample. This semi-supervised training strategy was proposed to take advantage of an abundant unlabeled corpus \cite{karita2019semi}. Our system can also benefit from this semi-supervised strategy, but we only focus on supervised training in this work. 

For the tied-layer loss, we calculated the symmetrized Kullback-Leibler divergence between the outputs of the text and speech encoders instead of the asymmetric one \cite{luong2019unified}:
\begin{multline}
\label{eq:losstie}
    \textrm{loss}_{tie} = \frac{1}{2} \; L_{KLD}(TEnc(\boldsymbol{x}), SEnc(\boldsymbol{y})) \\ 
    + \frac{1}{2}\; L_{KLD}(SEnc(\boldsymbol{y}), TEnc(\boldsymbol{x}))
\end{multline}
The constraints help obtaining a consistent latent space between the text and speech encoders. Through experiments we found that KL divergence is an effective tied-layer loss \cite{luong2019unified}\footnote{Karita \MakeLowercase{\textit{et al.}}\ \cite{karita2019semi} reported that KL divergence is unstable for training. The reason for this contrast is that in their work the autoencoder-based latent space is \textit{assumed} to be Gaussian distribution while in our case it is \textit{forced} to be an isotropic Gaussian distribution through VAE-like structure \cite{kingma2013auto}}.

As the text and speech encoders output the mean ($\mu$) and standard deviation ($\sigma$) of the LLE similarly to the VAE network, we need to apply the reparameterization trick so that the network can be trained with the backpropagation algorithm:
\begin{equation}
    \boldsymbol{z} = \boldsymbol{\mu} + \boldsymbol{\sigma} \odot \boldsymbol{\epsilon},\quad \boldsymbol{\epsilon} \sim \mathcal{N}(0,1) \; .
\end{equation}
The same process is used in the inference step to generate an LLE sequence. As $\epsilon$ is drawn from a normal distribution, this trick can also be interpreted as a noise augmenting process, which means the text and speech decoders are trained in a denoising fashion.
This, in turn, makes them robust to unseen samples, which is helpful for speaker adaptation.
To push the speech generation system toward an E2E setup, we include a neural vocoder to generate a waveform from the acoustic representation instead of using a conventional vocoder. In the training stage, the neural vocoder is trained separately from the rest of the system on natural speech samples:
\begin{equation}
    \textrm{loss}^\prime_{train} = \textrm{loss}_{voc}
\end{equation}
We used an autoregressive WaveNet \cite{van2016wavenet} conditioned on mel-spectrogram and trained on multi-speaker corpus as the neural vocoder in this paper. However, our voice cloning procedure is applicable to any type of neural vocoder.

\subsection{Speaker adaptation framework}
\label{sec:proposal}

The multimodal network trained in the previous stage is essentially a multi-speaker TTS/VC system; however our goal is to perform voice cloning for unseen speakers.
The following subsections describe the cloning protocols for the unsupervised scenario, which uses untranscribed speech, and the supervised scenario, which uses transcribed speech.

\subsubsection{Cloning voice using untranscribed speech}
\label{sssec:adaptation-process}
The core mechanism for unsupervised speaker adaptation is the same as our prior work \cite{luong2019unified,luong2019bootstrapping}; however, the detail of the executions have been updated. The voice cloning stage now contains three steps, which takes the neural vocoder into account.

\textbf{Step 1 - Adaptation:}
This is essentially our legacy unsupervised adaptation stage \cite{luong2019unified} in which the speech decoder and neural vocoder are adapted separately. We first remove all speaker components and then fine-tune the remaining parameters of the speech decoder using the following loss:
\begin{equation}
    \textrm{loss}_{adapt} = \textrm{loss}_{sts} + \beta \; \textrm{loss}_{cycle}
\end{equation}
The speech distortion $\textrm{loss}_{sts}$ by itself is enough for the adaptation \cite{luong2019unified}, but we further add a linguistic cycle consistent term $\textrm{loss}_{cycle}$ to try to improve the performance. $\textrm{loss}_{cycle}$ is the KL divergence between LLE distributions of natural speech and reconstructed speech as follows:
\begin{multline}
\label{eq:heatup}
    \textrm{loss}_{cycle} = \frac{1}{2} \; L_{KLD}(SEnc(\boldsymbol{y}), SEnc(\boldsymbol{\tilde{y}})) \\ 
    + \frac{1}{2}\; L_{KLD}(SEnc(\boldsymbol{\tilde{y}}), SEnc(\boldsymbol{y}))
\end{multline}
Even though both $\textrm{loss}_{sts}$ and $\textrm{loss}_{cycle}$ try to force the reconstructed features to be close to natural speech, they focus on different aspects; $\textrm{loss}_{sts}$ is either $l_1$ or $l_2$ frame-based hard distortion of the acoustic features, while $\textrm{loss}_{cycle}$ focuses on linguistic content with soft divergence.
We adapt the neural vocoder in a similar manner using its goal loss:
\begin{equation}
    \textrm{loss}^\prime_{adapt} = \textrm{loss}_{voc}
\end{equation}
As a neural vocoder depends on speech only, it can be used in an unsupervised adaptation strategy. This is a simple yet effective approach \cite{liu2018wavenet}.

\textbf{Step 2 - Welding:} 
Even though fine-tuning the acoustic model and the neural vocoder separately can produce sufficient quality \cite{liu2018wavenet}, there are still mismatches between the generated features and the natural features used to train the vocoder.
For text-to-speech systems, Zhao \MakeLowercase{\textit{et al.}}\ \cite{zhao2018wasserstein} fine-tuned an acoustic model with the losses propagating from a neural vocoder, while Ping \MakeLowercase{\textit{et al.}}\ \cite{ping2019clarinet} jointly trained them together. For voice conversion, due to the duration mismatch between source and target utterances, Huang \MakeLowercase{\textit{et al.}}\ proposed that the WaveNet vocoder be fine-tuned by using reconstructed acoustic features of a target speaker \cite{huang2019refined}.
Motivated by them, we deploy a ``welding'' strategy, illustrated in Fig. \ref{fig:step-welding}, that conducts fine-tuning by using the reconstructed features of the target speaker in a similar way to Huang's approach \cite{huang2019refined}, but, for both the speech decoder and neural vocoder like Ping's method \cite{ping2019clarinet} based on the loss function below:
\begin{equation}
    \textrm{loss}_{weld} = \textrm{loss}_{sts} + \gamma \; \textrm{loss}_{voc} \;,
\end{equation}
where $\textrm{loss}_{sts}$ is included to preserve the acoustic space even after the welding process as the speech decoder is assumed to be autoregressive in the domain.

Two practical tactics are further introduced for this step.
1) \textit{mean-value LLE}: to let the acoustic model learn fine-grained details, we remove the sampling process from the speech encoder and use the mean value instead.
2) \textit{mix-in}: as losses propagating from the neural vocoder can overpower the speech decoder \cite{zhao2018wasserstein}, we propose a mix-in tactic, inspired by drop-out, to ease this problem. Specifically the output of the speech decoder is randomly mixed with natural frames by a percentage to reduce the amount of losses propagated back and act as a form a regularization to prevent overfitting to the generated frames.

\textbf{Step 3 - Inference:}
Even though we use the speech encoder to tune the speech decoder and neural vocoder in the adaption and welding steps, the text encoder can utilize these tuned modules without any further adjustment in inference (See Fig.\ \ref{fig:step-inference}) thanks to the consistency between the latent spaces of the text and speech encoders. As our cloning method tunes entire modules, the more data available, the better the performance.

\begin{figure}[tb]
  \centering
  \subfloat[Step 1 - Adaptation (supervised alternative)\label{fig:stage-alternative}]{\includegraphics[width=0.7\linewidth]{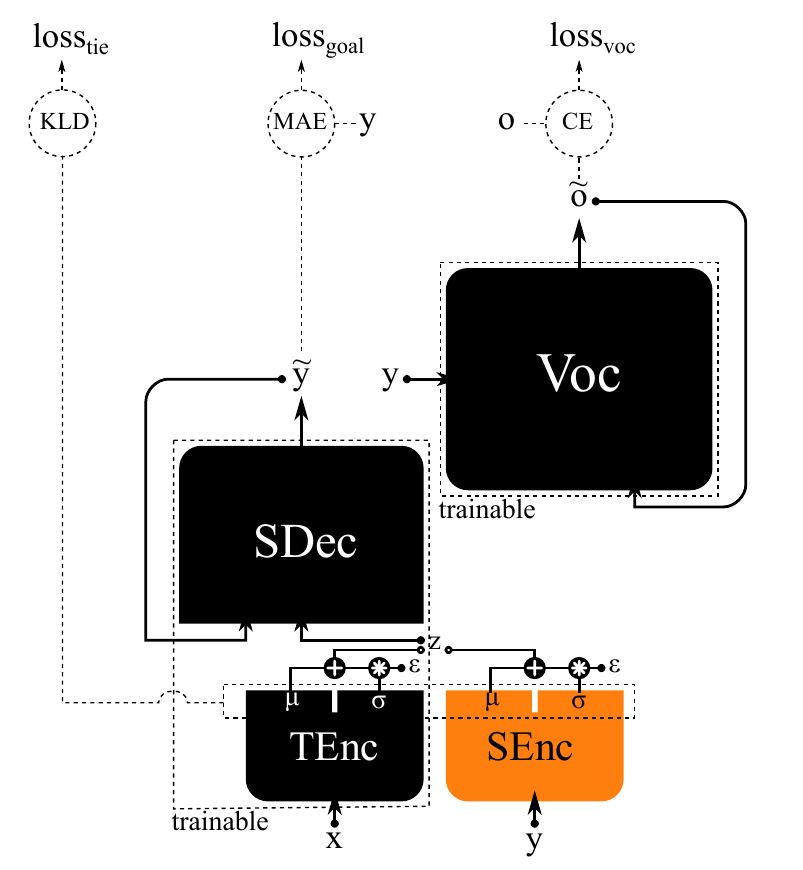}}
\caption{Cloning procedure with transcribed speech of a target speaker. The welding and inference steps are identical to the procedure with untranscribed speech. In this figure, $\textrm{loss}_{goal}$ is a placeholder for either $\textrm{loss}_{tts}$ or $\textrm{loss}_{sts}$ depending on the encoder/decoder combination.}
\vspace{-3mm}
\end{figure}

\subsubsection{Alternative strategy to cloning voices with transcribed speech}
\label{sssec:alternative-process}

The strategy for supervised speaker adaptation using transcribed speech was also refined compared with our previous work \cite{luong2019unified}.
Instead of using exactly the same strategy as those for the above unsupervised strategy, we first tune the speech decoder and text encoder together using the transcribed speech since transcriptions could benefit the TTS system. 

\textbf{Step 1 - Adaptation (supervised alternative):} The supervised strategy for the adaptation step is illustrated in Fig. \ref{fig:stage-alternative}. 
We adapt both the speech decoder and text encoder using the following function.
\begin{equation}
    \textrm{loss}^{\prime \prime}_{adapt} = \textrm{loss}_{tts} + \alpha \; \textrm{loss}_{sts} + \beta \; \textrm{loss}_{tie}
\end{equation}
The optimizing loss is similar to that used in the training stage (Equation \ref{eq:losstrain}). We use $\textrm{loss}_{sts}$ and $\textrm{loss}_{tie}$ to maintain the linguistic latent space for VC.
The \textit{welding} and \textit{inference} steps are the same as the unsupervised strategy.

\begin{figure*}[t]
\centering
  \begin{minipage}[t]{0.72\textwidth}
    \centering
    \subfloat[Text encoder][Text encoder\label{fig:blueprint-tenc}]{\includegraphics[width=0.33\linewidth]{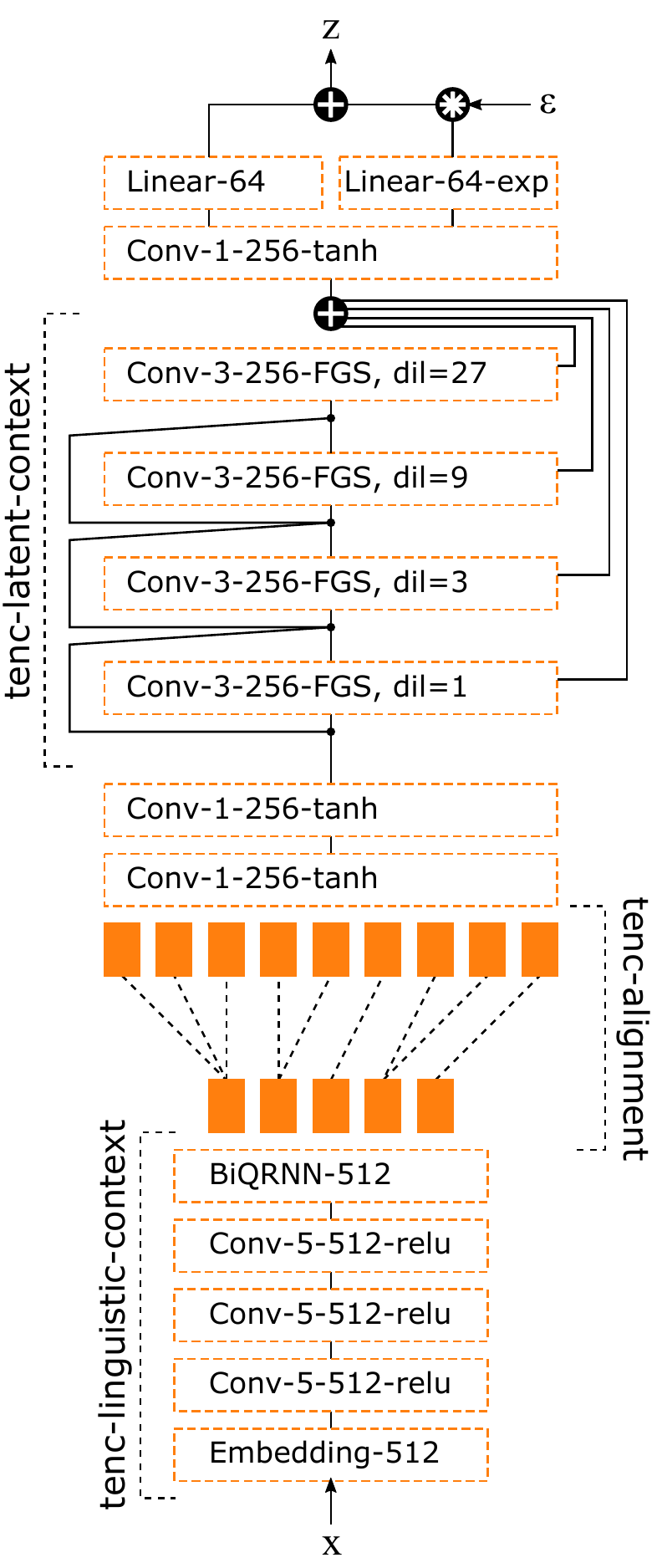}}
    \hfill
    \subfloat[Speech decoder][Speech decoder\label{fig:blueprint-sdec}]{\includegraphics[width=0.66\linewidth]{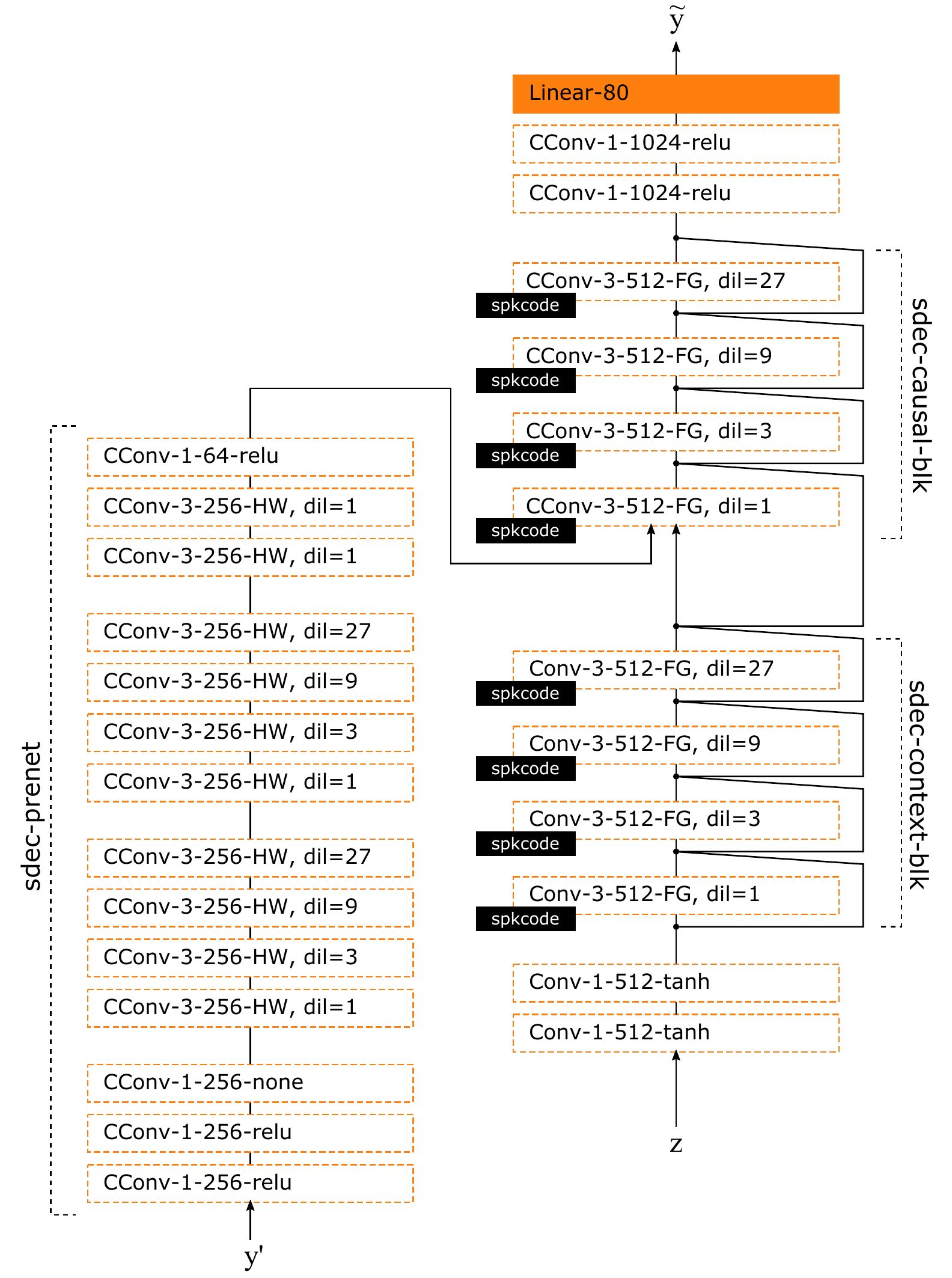}}
  \end{minipage}%
  \begin{minipage}[t]{0.24\textwidth}
    \vspace{4mm}
    \centering
    \subfloat[Text decoder][Text decoder\label{fig:blueprint-tdec}]{\includegraphics[width=0.99\linewidth]{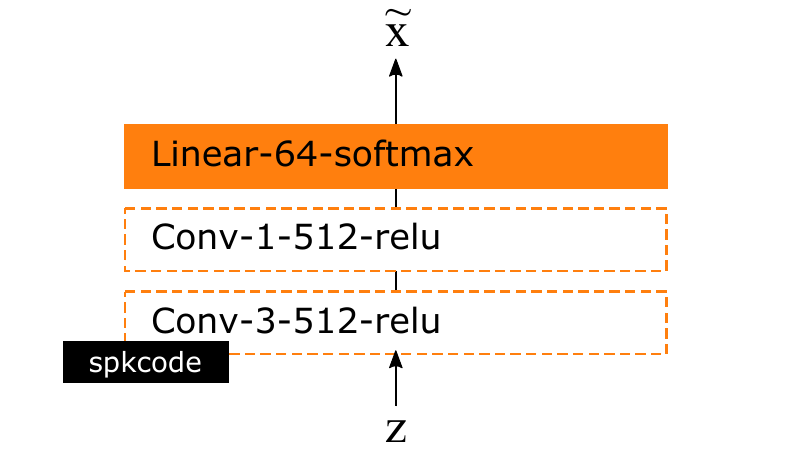}}\\
    \vspace{2.7mm}
    \subfloat[Speech encoder][Speech encoder\label{fig:blueprint-senc}]{\includegraphics[width=0.99\linewidth]{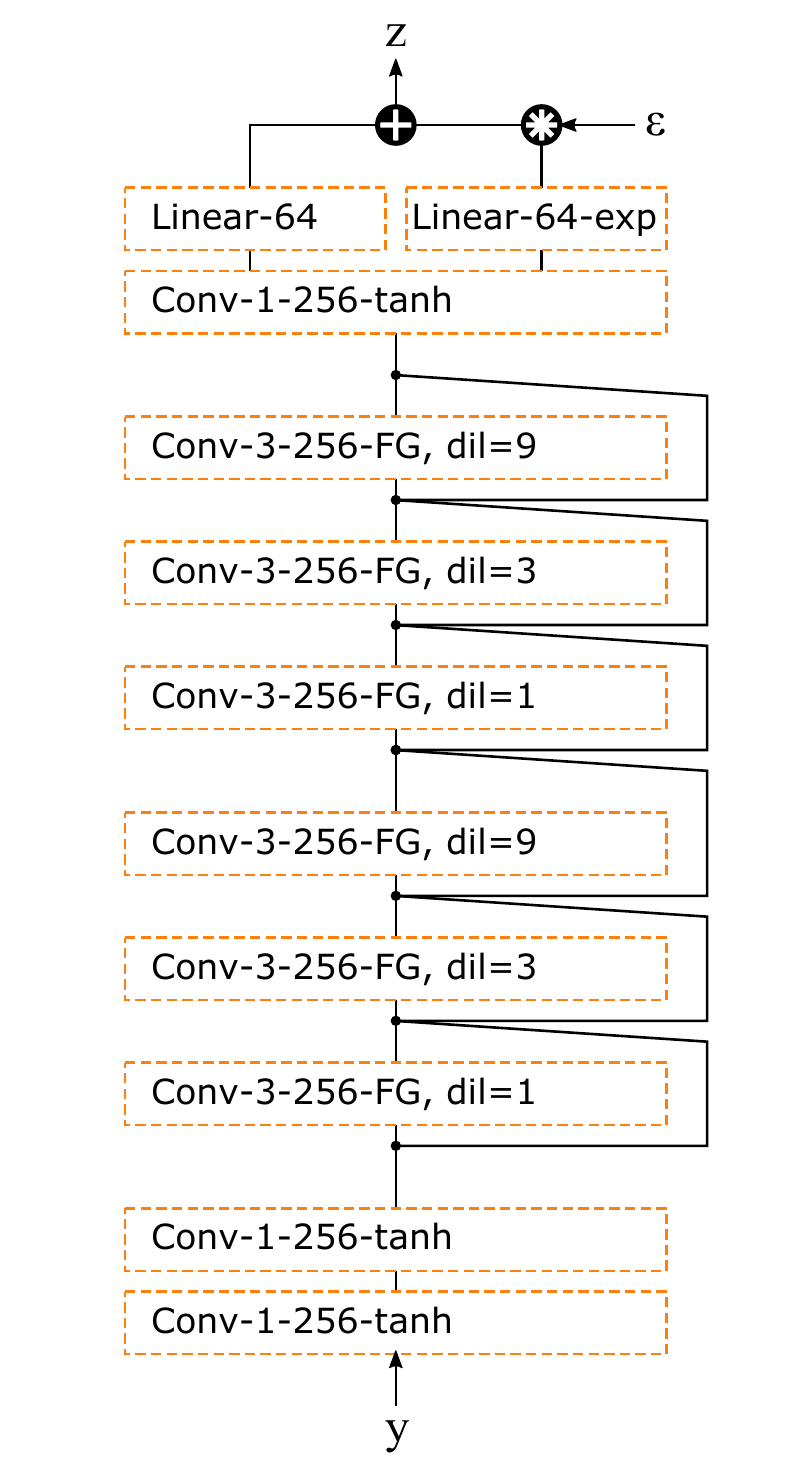}}
  \end{minipage}
  \caption{Blueprint of text-speech multimodal system. The naming convention is as follows \textit{type}-[\textit{filter}]-\textit{unit}-\textit{function}. Most layers are either causal (CConv) or non-causal (Conv) convolution layers with a \textit{filter} width of 3. Besides regular non-linear activation \textit{functions} like tanh or relu, we also use a non-linear filter-gate (FG), filter-gate with skip connection (FGS) and highway layer (HW). Dilation rate is indicated when applicable. The \textit{spkcode} black boxes indicate layers containing speaker bias components.}
  \vspace{-3mm}
\label{fig:blueprint}
\end{figure*}

\section{Details of NAUTILUS system}
\label{sec:condition}

The methodology explained in Sec.\ \ref{sec:system} can be applied to any neural architecture from the conventional acoustic model \cite{luong2019unified} to end-to-end (E2E) model \cite{shen2017natural}. Next we give the details on our system used in the experiments. It is not a fully E2E system but inspired by the E2E model in various ways. 

\subsection{Text-speech multimodal system}
Our system is shown in Fig. \ref{fig:blueprint}. The text representation $\boldsymbol{x}$ is a phoneme sequence and the speech representation $\boldsymbol{y}$ is mel-spectrogram.

\subsubsection{Text encoder}
the text encoder transforms a compact phoneme sequence $\boldsymbol{x}$ into the LLE sequence $\boldsymbol{z}$, which has the same length as the acoustic sequence.
Our specifications for the text encoder are illustrated in Fig. \ref{fig:blueprint-tenc}. The input phoneme sequence is represented as one-hot vectors.
As engineered linguistic features are no longer provided, \textit{tenc-linguistic-context} is used to learn the linguistic context. This is a direct imitation of Tacotron 2 \cite{shen2017natural} but with quasi-RNN \cite{bradbury2016quasi} used in place of the standard RNN to speed up the training.
The attention mechanism is essential in a E2E setup to unroll the phoneme sequence; our setup, however, uses an explicit duration/alignment module called ``\textit{tenc-alignment}'' in training and inference to have direct control over the generated sample prosody.\footnote{The \textit{tenc-alignment} could be replaced with attention mechanism for convenience, and this could also potentially improve the quality further \cite{watts2019improvements}.} 
The coarse linguistic features, then, go through several dilated convolution layers called ``\textit{tenc-latent-context}'' to capture the local context and smooth out the coarseness. 
\textit{tenc-latent-context} has essentially the same design as the acoustic models used in our prior work \cite{luong2019unified}, which used residual, skip connection and filter-gate function (Fig. 4a in \cite{luong2019unified}) to help the gradient flow:
\begin{equation}
    \label{eq:layerfg}
    \boldsymbol{h}_{l} = \tanh( \boldsymbol{W}^{f}_{l} \boldsymbol{h}_{l-1} + \boldsymbol{c}^f_l) \odot \sigma( \boldsymbol{W}^{g}_{l} \boldsymbol{h}_{l-1} + \boldsymbol{c}^g_l) \;,
\end{equation}
where $\boldsymbol{h}_{l}$ is the output of the $l$-th layer, and $\boldsymbol{W}^{f}_{l}$, $\boldsymbol{W}^{g}_{l}$, $\boldsymbol{c}^f_l$, and $\boldsymbol{c}^f_l$ are the weights and biases for filters and gates. The output of the text encoder consists of the mean and standard deviation of a text-encoded LLE sequence.

\subsubsection{Speech decoder} 
the speech decoder takes in an LLE sequence $\boldsymbol{z}$ to generate a respective acoustic sequence $\widetilde{\boldsymbol{y}}$ with a particular voice. It is essentially a multi-speaker speech synthesis model and there are three  components that significantly affect the performance: temporal context capturing \cite{zen2015unidirectional}, autoregressive mechanism \cite{wang2017autoregressive,watts2019improvements}, and speaker modeling \cite{luong2018scaling}.
\textit{sdec-context-blk} captures LLE temporal context by using time-domain convolution layers, which also contain speaker biases in their filters and gates (Fig. 4b in \cite{luong2019unified}):
\begin{multline}
    \label{eq:layerfgs}
    \boldsymbol{h}_{l} = \tanh( \boldsymbol{W}^{f}_{l} \boldsymbol{h}_{l-1} + \boldsymbol{c}^f_l +\boldsymbol{b}_l^{f,(k)} )  \\
    \odot \sigma( \boldsymbol{W}^{g}_{l} \boldsymbol{h}_{l-1} + \boldsymbol{c}^g_l + \boldsymbol{b}_l^{g,(k)} ) \;,
\end{multline}
where $\boldsymbol{b}_l^{f,(k)}$ and $\boldsymbol{b}_l^{f,(k)}$ are the speaker biases of $k$-th speaker in the training speaker pool. 
The effective type of speaker component depends on the network structure as well as the acoustic features \cite{luong2018scaling}. We previously found that speaker biases work the best for our setup \cite{luong2019unified}.

An autoregressive mechanism is introduced to improve the overall naturalness. \textit{sdec-prenet} is responsible for the autoregressive dependency that captures the past outputs using causal layers. This is a direct imitation of the \textit{AudioEnc} proposed by Tachinaba \MakeLowercase{\textit{et al.}}\ \cite{tachibana2018efficiently}. 
The layers in \textit{sdec-prenet} use the highway function in the same way as \cite{tachibana2018efficiently} as follows:
\begin{align}
    \label{eq:layerhw}
    \boldsymbol{h}_{l}^{f} &= \boldsymbol{W}^{f}_{l} \boldsymbol{h}_{l-1}, \\
    \boldsymbol{h}_{l}^{g} &= \sigma( \boldsymbol{W}^{g}_{l} \boldsymbol{h}_{l-1} ), \\
    \boldsymbol{h}_{l} &= \boldsymbol{h}_{l}^{f} \odot \boldsymbol{h}_{l}^{g} +  \boldsymbol{h}_{l-1} \odot (1-\boldsymbol{h}_{l}^{g})
\end{align}
The linguistic context and the past-state token are fed into more causal layers before being transformed into the acoustic features. The architecture of the speech decoder is shown in Fig. \ref{fig:blueprint-sdec}. 
We use the mean absolute error (MAE) as the loss function for the speech generation goals. In the adaptation stage, speaker biases are removed from the speech decoder.

\subsubsection{Speech encoder} 
the speech encoder extracts the LLE $\boldsymbol{z}$ from a given acoustic sequence $\boldsymbol{y}$ while stripping unnecessary information (i.e. speaker characteristics).
It is similar to an ASR model as the output needs to be independent from training speakers, and the model needs to be generalized to unseen targets.
We have no strong preference for speech encoder specification and simply use several dilated layers to capture the local context as illustrated in Fig. \ref{fig:blueprint-senc}. 

\subsubsection{Text decoder}
the text decoder takes an LLE sequence $\boldsymbol{z}$ and predicts the phoneme posterior $\widetilde{\boldsymbol{x}}$ at each frame. This is a new component introduced in this work compared with previous ones \cite{luong2019unified}. Unlike other modules that would be reused in various stages, the shallow text decoder is included in the training only and acts as an auxiliary regularizer.
Its purpose is forcing the latent linguistic embedding to focus more on phoneme information, which we found important for generating utterances with clear pronunciation. The balance between phoneme and other linguistic information is adjustable using the joint-goal weight $\alpha_{stt}$ and the representative power of the text decoder itself. This is why we use a couple of layers only to model the text decoder (Fig. \ref{fig:blueprint-tdec}). The cross-entropy criterion is used as the loss function of the phoneme classifier.

\subsection{WaveNet vocoder}
An auto-regressive WaveNet model conditioned on a mel-spectrogram \cite{hayashi2017investigation,shen2017natural,liu2018wavenet} is used as the neural vocoder of our setup. WaveNet is trained on either 22.05kHz or 24kHz speech depending on the scenarios. Waveform amplitudes are quantized by using 10-bit $\mu$-law. The network consists of 40 dilated causal layers containing speaker biases. Both the residual and skip channels are set at 128. This is a typical setup for WaveNet \cite{van2016wavenet}.  In the adaptation stage, speaker biases are removed before fine-tuning. 

\subsection{Training, adapting, and inferring configurations}

The General American English lexicon \cite{richmond2009robust} was used for text representation, and 56 distinct phonemes were found in our training data. An 80-dimensional mel-spectrogram was used as acoustic representation. The mel-spectrogram was calculated by using a 50-ms window size and 12.5-ms shift size. This was inspired by the setup of E2E TTS model \cite{shen2017natural,tachibana2018efficiently}.
The weighting parameters of the optimizing losses were $\alpha=0.1$, $\beta=0.25$ and $\gamma=0.01$. The learning rate was set at 0.1 for all optimizing stages. 
The dropout rate was set at 0.2 for most components apart from \textit{tenc-linguistic-context} and \textit{sdec-prenet}, for which the rate was set at 0.5.
The training was stopped when loss on validation stopped improving for ten consecutive epochs.

One hundred speakers of the VCTK corpus \cite{veaux2017superseded} were used to train the multi-speaker text-speech system and the WaveNet vocoder. The sampling rate was converted to the target scenarios.
Among the remaining speakers, one male and one female with an American accent were used as targets for an experiment described in Sec. \ref{ssec:exp2}. All common sentences were removed from the training so they could be used for evaluation.
As VCTK lacks diversity in linguistic content, we first used 24-kHz LibriTTS corpus \cite{zen2019libritts} to warm-up the text-speech network.
Only \textit{train-clean-100} and \textit{train-clean-360} sets, which are 245 hours in total, were used to reduce the warming time.
The phoneme alignments of each corpus were extracted using an ASR model trained on the same corpus using the KALDI toolkit \cite{povey2011kaldi}.
For evaluated utterances, the model trained on the LibriTTS corpus is used to extract their phoneme alignments.

There were two voice cloning experiments, scenarios A and B. For the voice cloning stage, the number of epochs was fixed to create a uniform process.
Specifically, for scenario A described in Sec. \ref{ssec:exp1}, we first adapted the text-speech model for 256 epochs, the vocoder for 128 epochs, and then welded them together for 64 more. For scenario B described in Sec. \ref{ssec:exp2}, the number of epochs was 256, 64, and 32, respectively. 
The \textit{mix-in} rate in the welding step was set at 0.9.

For the inference stage, the speech encoder used its mean output for VC while text encoder sampled a LLE sequence from Gaussian distributions for TTS as shown in Fig. \ref{fig:step-inference}.
To maintain stochasticity but reduce the chance of sampling undesirable outliers, we multiplied the standard deviation output of the text encoder by 0.1 before random sampling.

\subsection{Evaluation measurements}

We treated our system as a whole, instead of focusing on individual techniques, and we compared it with other third-party systems. 
For objective evaluations, we used an ASR model\footnote{A chain system based on TDNN-F pretrained on the Librispeech corpus \cite{panayotov2015librispeech} was used for the calculation (\url{http://kaldi-asr.org/models/m13}).} to calculate the word error rate (WER) of the generated speech.
Note that WER should be treated as a secondary reference since it is highly sensitive to the training data of the ASR model. As a large-scale English corpus of native speakers was used to train the speech recognition model, we can interpret lower WER as indicating better pronunciation and/or greater similarity to the voices of speakers in the training set.
For subjective evaluations, we used MOS on a 5-point scale for quality and DMOS on a 4-point scale for speaker similarity \cite{lorenzo2018voice}. In most of the questions on speaker similarity, participants were asked to compare the speaker similarity of a generated utterance with a natural utterance. However, scenario A included additional questions for comparing speaker similarity between generated utterances. In scenario B, the participants were also asked to do several AB tests on quality and speaker similarity. In the AB test, two speech samples were shown at each test page and participants were asked to choose the better of the two. These questions were used to highlight the fine-grained differences between generation systems. Each participant in our subjective listening tests was asked to do ten sessions.

\begin{table}[tb]
    \caption{Target speakers of scenario A}
    \centering
    \scalebox{0.95}{
    \begin{tabular}{llllrr}
        \hline \hline
        Speaker & Database & Gender & Accent & Quantity & Duration \\\hline
        VCC2TF1 & VCC2018 & female & American & 81 utt. & 5.2 min \\
        VCC2TF2 & VCC2018 & female & American & 81 utt. & 5.0 min \\
        VCC2TM1 & VCC2018 & male & American & 81 utt. & 5.2 min \\
        VCC2TM2 & VCC2018 & male & American & 81 utt. & 5.3 min\\ \hline
    \end{tabular}}
    \label{tab:exp1-data}
\vspace{-3mm}
\end{table}

\begin{table}[tb]
    \caption{Word error rate for objective evaluation of scenario A }
    \centering
    \scalebox{0.95}{
    \begin{tabular}{lrrrr}
        \hline \hline
        System &  \multicolumn{4}{c}{Target speakers (\%WER)} \\
         &  VCC2TF1 & VCC2TF2 & VCC2TM1 & VCC2TM2 \\\hline
        XV & 3.25 & 2.98 & 3.66 & 10.57  \\ \hline
        N10$^=$ & 9.21 & 7.99 & 11.79 & 9.89  \\
        N10$^\times$ & 9.62 & 11.52 & 8.67 & 9.21 \\ \hline
        N13$^=$ & 23.31 & 21.68 & 31.57 & 27.37 \\ 
        N13$^\times$ & 32.25 & 24.80 & 21.41 & 26.96 \\
        N17$^=$ & 25.47 & 24.39 & 33.47 & 23.71 \\ 
        N17$^\times$ & 38.08 & 31.44 & 35.23 & 25.88 \\ \hline
        VCA$_u^=$ & 25.34 & 26.02 & 27.37 & 25.75 \\ 
        VCA$_u^\times$ & 30.62 & 27.51 & 23.71 & 22.63 \\ \hline
        TTS$_u$ & 7.72 & 8.40 & 6.23 & 7.18 \\
         &  \multicolumn{4}{c}{Source speakers (\%WER)} \\
         &  VCC2SF3 & VCC2SF4 & VCC2SM3 & VCC2SM4 \\\hline
        S00 & 5.69 & 4.88 & 5.69 & 7.32  \\ \hline
    \end{tabular}}
    \label{tab:exp1-objective}
\vspace{-3mm}
\end{table}

\section{Experiment scenarios and evaluations}
\label{sec:scenarios}

As our system can clone voices by using either transcribed or untranscribed speech and can be used as TTS or VC systems, it would be difficult to evaluate all of these tasks in a single experiment. Therefore, we tested its performance and versatility under two separate scenarios.
The first scenario focuses more on VC and cloning voices with untranscribed speech, while the second scenario focuses more on TTS and performance of the supervised and unsupervised speaker adaptation strategies\footnote{The generated speech samples of both experiment scenarios are available at \url{https://nii-yamagishilab.github.io/sample-versatile-voice-cloning/}}.

\subsection{Cloning voices using untranscribed speech}
\label{ssec:exp1}

In the first scenario, scenario A, we tested the ability to clone voices by using a small amount of untranscribed speech (about five minutes).
A system showing good performance under this scenario is expected to have the capability to clone thousands of voices efficiently and cheaply.

\begin{figure}[tb]
  \centering
  \includegraphics[width=0.95\columnwidth]{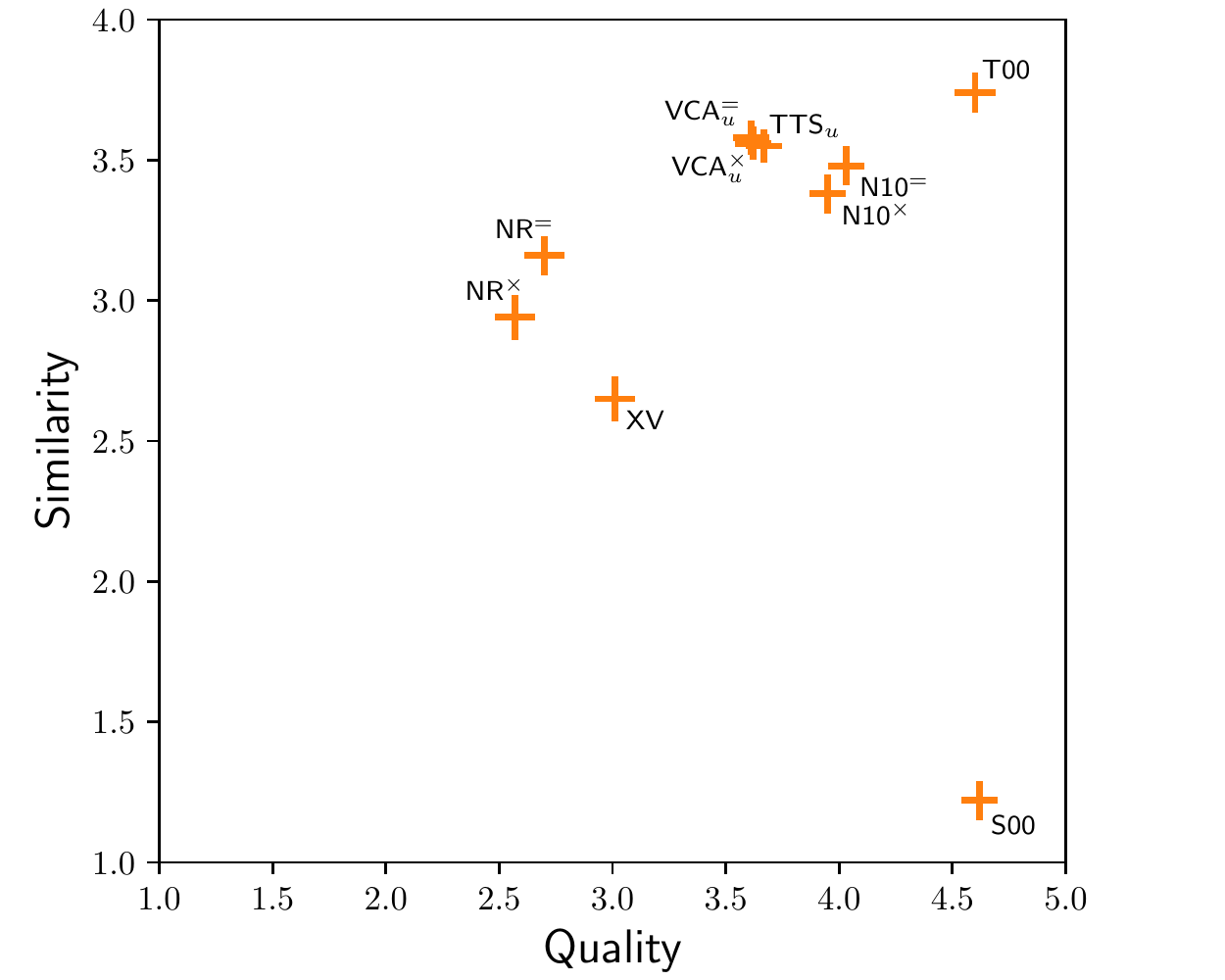}\\
  \includegraphics[width=0.95\columnwidth]{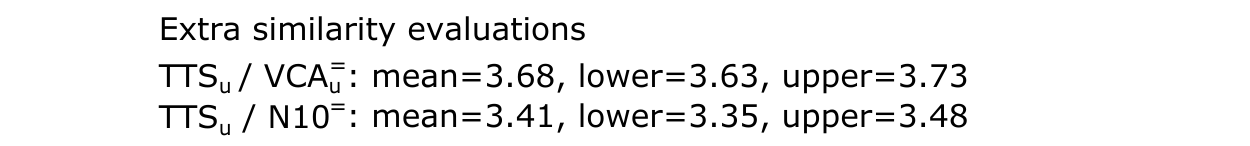}\\
  \vspace{-2mm}
\caption{Subjective results of scenario A. Lines indicate 95\% confidence interval. Cross-gender and Same-gender conversion of VC systems were treated as separate entities.}
\vspace{-5mm}
\label{fig:exp1-subjective}
\end{figure}

\subsubsection{Experiment setups}
we re-enacted the SPOKE task of Voice Conversion Challenge 2018 (VCC2018) \cite{lorenzo2018voice} for this scenario. 
The original goal of the task was to build VC systems for 4 target English speakers (2 males and 2 females) using 81 utterances (Table \ref{tab:exp1-data}). These systems were used to convert the speech of 4 source speakers (2 males and 2 females) into each of the target voices. 
We followed the VCC2018 guideline \cite{lorenzo2018voice} faithfully with one extension -- we evaluated TTS systems as well as VC systems at the same time. These TTS systems were required to train on the untranscribed speech of the target speakers. In the inference stage, transcriptions of source utterances were used to generate speech with TTS systems.
As there were only 35 unique sentences, we generated each sentence twice.
In summary, each TTS system produced 70 utterances for each target speaker while each VC system produced 140 utterance. 
We split each VC system into two entities, one for same-gender conversion denoted by the superscript ``$=$'' and the other for cross-gender denoted by ``$\times$''.

\subsubsection{Evaluated systems}

We evaluated the following TTS and VC systems in scenario A: 
\begin{itemize}
  \item \textbf{XV}: a speaker-adaptive E2E TTS system using the x-vector \cite{arik2018neural,jia2018transfer,cooper2019zero}. XV was used as a third-party unsupervised TTS baseline. We used the \textit{libritts.tacotron2.v1} model and the speaker-independent WaveNet vocoder \textit{libritts.wavenet.mol.v1} which were trained on the LibriTTS corpus to realize this approach. Both are available at the ESPnet \cite{watanabe2018espnet} repository\footnote{https://github.com/espnet/espnet}. As the x-vector is utterance-based, we randomly picked five utterances (about ten seconds) from the training pool of target speakers to extract the x-vector each time we generate an utterance.

  \item \textbf{N10}: the winner of the VCC2018 SPOKE task. N10 contains a PPG-based acoustic model \cite{sun2016phonetic} and a fine-tuned WaveNet vocoder \cite{liu2018wavenet}. It uses a speaker-independent ASR model trained on hundreds of hours of labeled data to extract PPG from speech. N10 clones voice without using the speech data of source speakers.
  
  \item \textbf{N13}\textbackslash \textbf{N17} (\textbf{NR}): the runners-up of the VCC2018 SPOKE task in terms of quality and similarity, respectively. To reduce the amount of systems, we treat them as one (denotes as NR) and use N13 in the quality evaluation while using N17 \cite{wu2018nu} in the similarity evaluation. 
  
  \item \textbf{VCA}$_\text{u}$: VC mode of the NAUTILUS system which was adapted to target speaker by using the unsupervised strategy described in Sec.\ \ref{sssec:adaptation-process}. The letter ``\textit{A}'' as in ``\textit{any-to-one}'' indicates that the model is not trained on source speakers. The word \textit{unsupervised} means that the cloning is performed with untranscribed speech in the context of our current work. It is operated at 22.05 kHz to be compatible with the target speakers.

  \item \textbf{TTS}$_\text{u}$: TTS mode of the NAUTILUS system which was adapted by using the unsupervised strategy. As we did not train an automatic duration model, we used the duration extracted from the same-gender source speakers to generate speech from text. This means that TTS$_\text{u}$ shares the same duration model as VCA$_\text{u}^{=}$ (and other same-gender VC systems). This reduces the difference in experimental conditions between them and allows us to make more insightful observations.
  
  \item \textbf{T00} and \textbf{S00}: natural utterances of the target and source speakers used as references, respectively.
\end{itemize}

\begin{table}[tb]
    \caption{Target speakers of scenario B}
    \centering
    \scalebox{0.95}{
    \begin{tabular}{llllrr}
        \hline \hline
        Speaker & Database & Gender & Accent/L1 & Quantity & Duration \\\hline
        p294 & VCTK & female & American & 325 utt. & 11.2 min \\
        p345 & VCTK & male & American & 325 utt. & 11.0 min \\
        MF6 & EMIME & female & Mandarin & 145 utt. & 10.2 min \\
        MM6 & EMIME &male & Mandarin & 145 utt. & 11.3 min\\ \hline
    \end{tabular}}
    \label{tab:exp2-data}
\vspace{-3mm}
\end{table}

\subsubsection{Evaluation} 
Twenty-eight native English speakers participated in the subjective test for scenario A. They were asked to answer 18 quality and 22 similarity questions in each session. In summary, each system was judged 560 times for each measurement, while natural speech systems (T00 and S00) were judged 280 times.
The objective and subjective evaluation results are shown in Table \ref{tab:exp1-objective} and Fig. \ref{fig:exp1-subjective} with many interesting observations. a) XV had better quality but worse similarity than the runners-up of VCC2018, while it received the lowest WER for certain speakers. One possible explanation is that the utterances generated by XV had the characteristics of the speakers in LibriTTS corpus instead of those of the target speakers, which makes its utterances more compatible with ASR model trained on LibriSpeech. The subjective evaluation of the XV speech samples supports this speculation. b) Our systems had high scores in both subjective measurements. Interestingly our TTS system had a lower WER than our VC systems. c) Even though we had a lower score for quality than did N10, the similarity seemed to be higher. d) Our TTS and VC systems had highly consistent results, while there was a gap between the same-gender and cross-gender subsystems of N10.
The extra similarity evaluations, between the generated systems, presented in Fig.\ \ref{fig:exp1-subjective}, shows similar results. 
The similarity between our TTS$_u$ and VCA$_u^=$ systems was higher than the similarity between TTS$_u$ and N10$^=$.

\subsubsection{Scenario conclusion} Even though the naturalness of our voice cloning system was slightly worse than N10 (again the best system at VCC2018), generally speaking it has achieved performance that is comparable to SOTA systems considering the difference in experimental conditions (e.g., the amount of data used in the training stage). More importantly, our system can seamlessly switch between TTS and VC modes with high consistency in terms of speaker characteristics. This is a desirable trait that would be useful for many applications.

\subsection{Capturing unique speaker characteristics}
\label{ssec:exp2}

\begin{table}[tb]
    \caption{Word error rate for objective evaluation of scenario B}
    \scalebox{0.95}{
    \begin{tabular}{lrrrr}
        \hline \hline
        System &  \multicolumn{4}{c}{Target speakers (\%WER)} \\
         &  VCTK-p294 & VCTK-p345 & EMIME-MF6 & EMIME-MM6 \\\hline
        NAT* & 6.09 & 8.69 & 56.24 & 43.39  \\ \hline
        XV & 3.50 & 24.05 & 5.33 & 3.81  \\ 
        FT & 13.39 & 20.09 & 57.53 & 42.01  \\ \hline
        VCM$_u$ & 22.22 & 24.05 & 27.70 & 27.09  \\ 
        VCM$_s$ & 23.29 & 24.81 & 29.07 & 29.53  \\ \hline
        TTS$_u$ & 8.37 & 9.74 & 13.39 & 14.92  \\ 
        TTS$_s$ & 9.28 & 10.05 & 36.38 & 38.20  \\
         &  \multicolumn{4}{c}{Source speakers (\%WER)} \\
         &  VCTK-p299 & VCTK-p311 & - & - \\\hline
        SRC** & 5.64 & 6.51 & - & - \\ \hline
    \end{tabular}}
    \label{tab:exp2-objective}
\vspace{1mm}
*calculated on all training utterances of target speakers.

**calculated on natural utterances of source speakers.
\vspace{-3mm}
\end{table}

\begin{figure*}[t]
  \captionsetup[subfigure]{justification=centering}

  \subfloat[Native speakers as targets\label{fig:exp2-subjective-native}]{
    \begin{minipage}[t]{0.32\textwidth}
    \centering
    \includegraphics[width=1.0\linewidth]{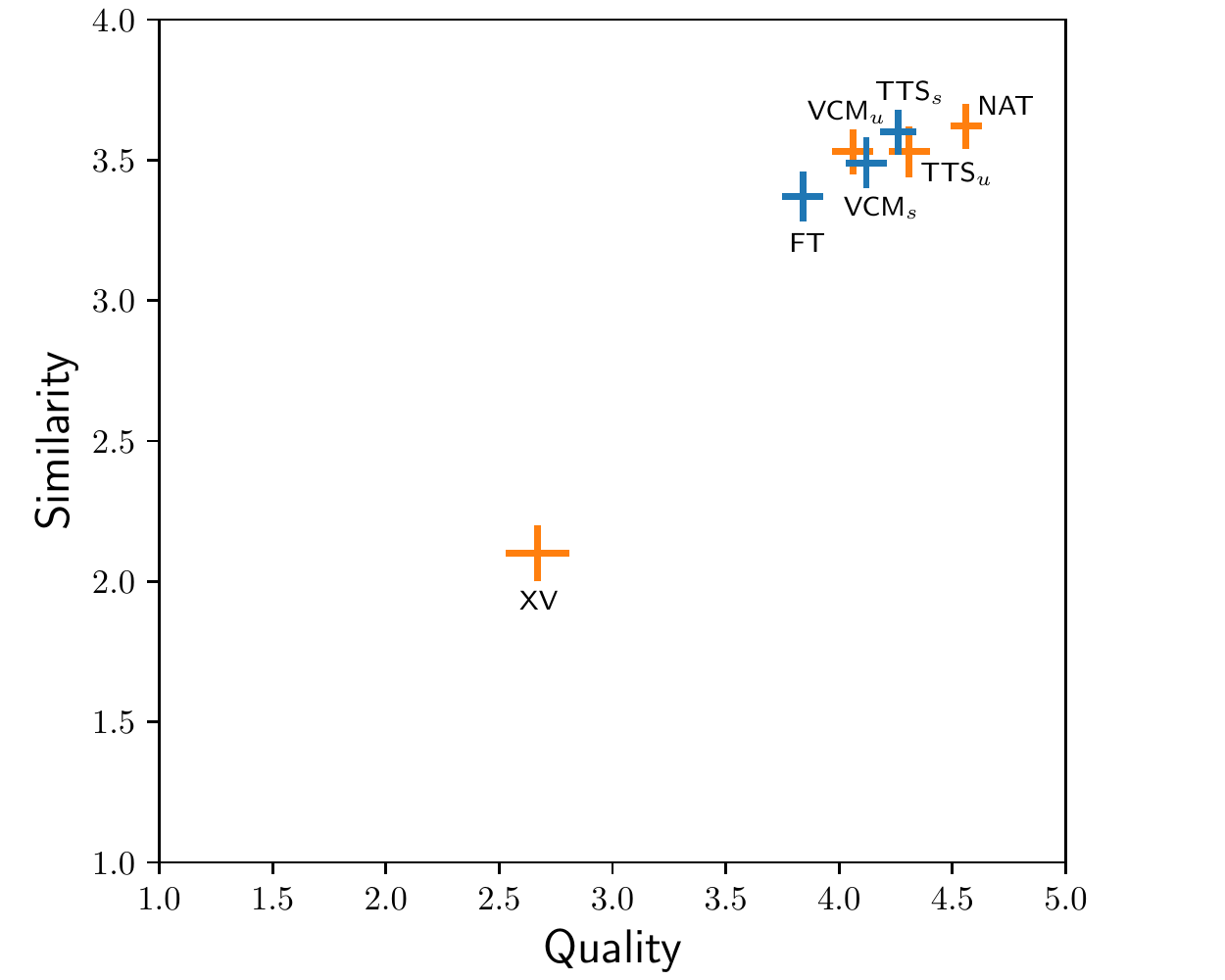} \\
    \vspace{2mm}
    \includegraphics[width=1.0\linewidth]{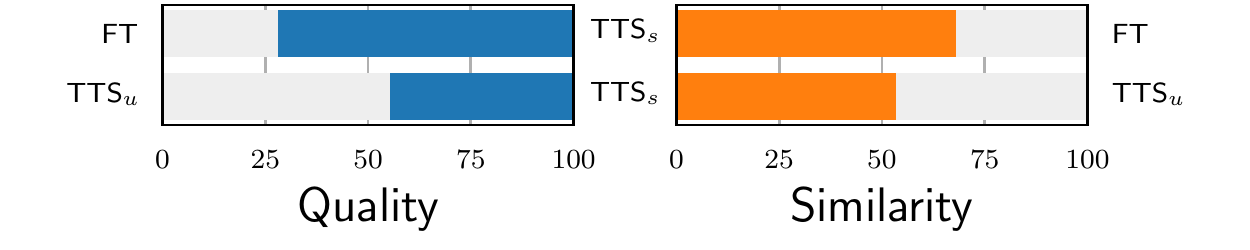}
    \end{minipage}
  }
  \subfloat[Non-native speakers as targets\label{fig:exp2-subjective-nonnative}]{
    \begin{minipage}[t]{0.32\textwidth}
    \centering
    \includegraphics[width=1.0\linewidth]{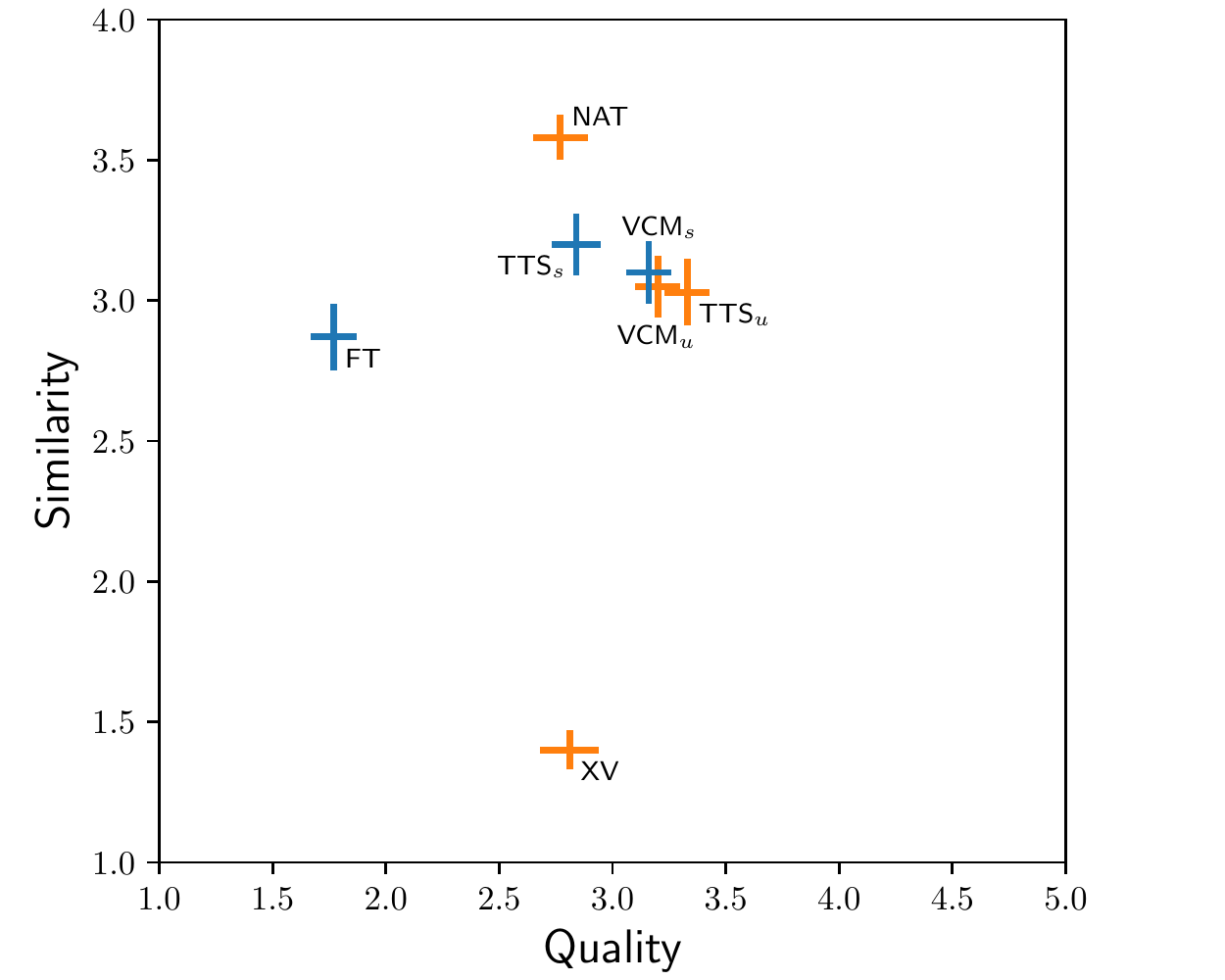} \\
    \vspace{2mm}
    \includegraphics[width=1.0\linewidth]{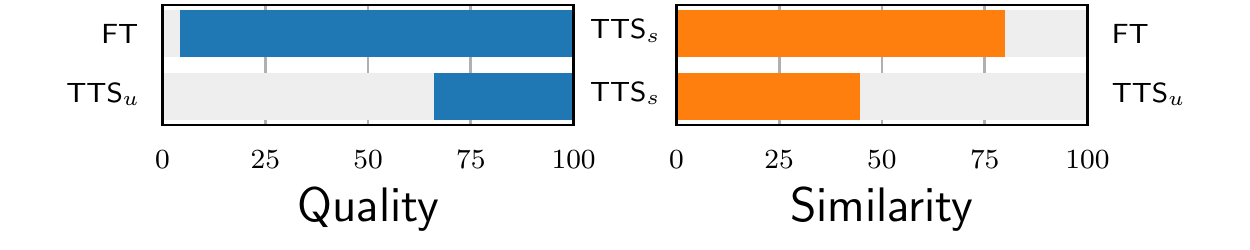}
    \end{minipage}
  }
  \subfloat[Per-listener non-native NAT][Details per listener evaluation of \\non-native natural utterances (NAT)\label{fig:exp2-subjective-nat}]{
    \begin{minipage}[t]{0.32\textwidth}
      \includegraphics[width=1.0\linewidth]{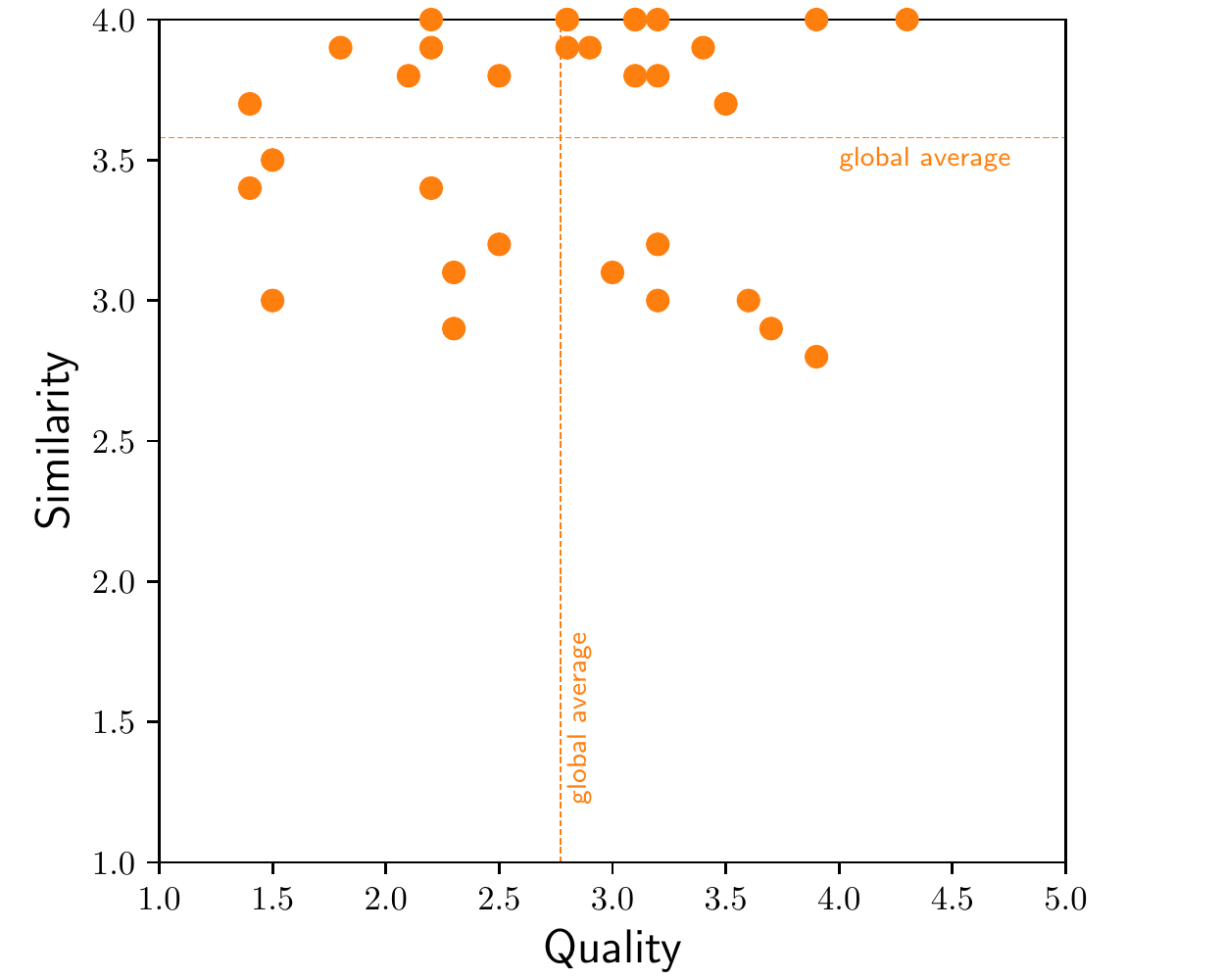}
    \end{minipage}
  }
\caption{Subjective evaluations of scenario B. The lines used to form the crosses in figure (a) and (b) indicate 95\% confidence interval.}
\vspace{-4mm}
\label{fig:exp2-subjective}
\end{figure*}

As mentioned earlier, the way voice cloning is differentiated from speech synthesis is that it should prioritize capturing the unique characteristics of target speakers.
While it is easy for listeners to grasp general global characteristics (e.g., average pitch), it is more difficult to notice local subtle traits (e.g., pronunciation of particular words) with just a single reference utterance.
We could use famous individuals as targets \cite{lorenzo2018can}, but this assumes that listeners would be familiar with them. 
In scenario B, we therefore used non-native speakers as targets to highlight their unique characteristics. This is convenient for subjective evaluation as native speakers can generally spot their distinctiveness without any explanation about the linguistic aspect of it \cite{janska2010native}. 
In simple words, the goal of scenario B was to reproduce the accent of non-native speakers. This scenario is closely related to reducing accents \cite{aryal2014can,oshima2016non} or controlling accents \cite{zhang2019learning} tasks.

\subsubsection{Experiment setups}
the target speakers for this scenario included two American and two non-native English speakers who use Mandarin as their native language. Each speaker had about 10 minutes of speech as listed in Table \ref{tab:exp2-data}. As the base model was trained with native speakers of English, the speakers from the VCTK corpus represented the standard easy task while the speakers from the EMIME corpus \cite{wester2011emime} represented difficult and unique target speakers.
The evaluated systems were required to be built with either the transcribed or untranscribed speech of the targets.
Twenty common sentences from the VCTK corpus were used for the evaluations. Each sentence was generated twice by each TTS system, which totaled 40 utterances. In the case of VC, one female (p299) and one male (p311) with a general American accent included in the training pool are used as source speakers.

\subsubsection{Evaluated systems}

The following TTS and VC systems were used for the evaluation in scenario B:
\begin{itemize}

  \item \textbf{XV}: the same x-vector system in scenario A is reused as the unsupervised baseline of TTS.
  
  \item \textbf{FT}: a fine-tuned E2E TTS system was used as the supervised baseline. We used \textit{ljspeech.tacotron2.v3}, implemented with ESPnet \cite{hayashi2020espnet}, as the initial model. It was trained with 24 hours of the transcribed speech of a female speaker from the LJSpeech corpus \cite{ito2017lj}. 
  An initial WaveNet vocoder was also trained with the same corpus.
  When cloning voices, we fine-tuned both acoustic and vocoder models with the transcribed speech of the targets.
  This system represented a simple supervised approach by fine-tuning a well-trained single speaker model \cite{inoue2020semi}.
  
  \item \textbf{VCM}$_\text{u}$: VC mode of the NAUTILUS system which was adapted to target speaker by using the unsupervised strategy described in Sec.\ \ref{sssec:adaptation-process}. The letter ``M'' as in ``many-to-one'' indicates that the source speakers were included in the training pool of the base model. The system was operated in 24kHz.
  
  \item \textbf{VCM}$_\text{s}$: VC mode of the NAUTILUS system which was adapted to target speaker by using the alternative supervised strategy described in Sec.\ \ref{sssec:alternative-process}. The supervised strategy is expected to be more relevant to TTS, but we included its VC counterpart as an anchor for comparison.
  
  \item \textbf{TTS}$_\text{u}$: TTS mode of the NAUTILUS system which was adapted by using the unsupervised strategy. The duration was extracted from the source speakers of VC. This means our TTS and VC systems share the same duration model.
  
  \item \textbf{TTS}$_\text{s}$: TTS mode of the NAUTILUS system which was adapted by using the alternative supervised strategy.
  
  \item \textbf{NAT}: the natural utterances of the target speakers.
\end{itemize}

\subsubsection{Evaluation}

Thirty-two native speakers took part in our subjective evaluation for scenario B.
As the participants were native English speakers living in Japan and many work as English teachers, we expected that they could quickly pick up on the non-native accents.
Each session had 18 quality and 18 similarity questions that contain utterances of both native and non-native speakers. Besides the standard MOS tests, we also included several AB tests in this scenario.
In summary, each system was evaluated 640 times for each assessment. The objective evaluation result are listed in Table \ref{tab:exp2-data}, and the subjective evaluation results are shown in Fig. \ref{fig:exp2-subjective}. Here the results of native and non-native speakers are separately shown.

\begin{figure*}[tb]
  \centering
  \subfloat[Speech generation goals\label{fig:loss-mae}]{\includegraphics[width=0.32\linewidth]{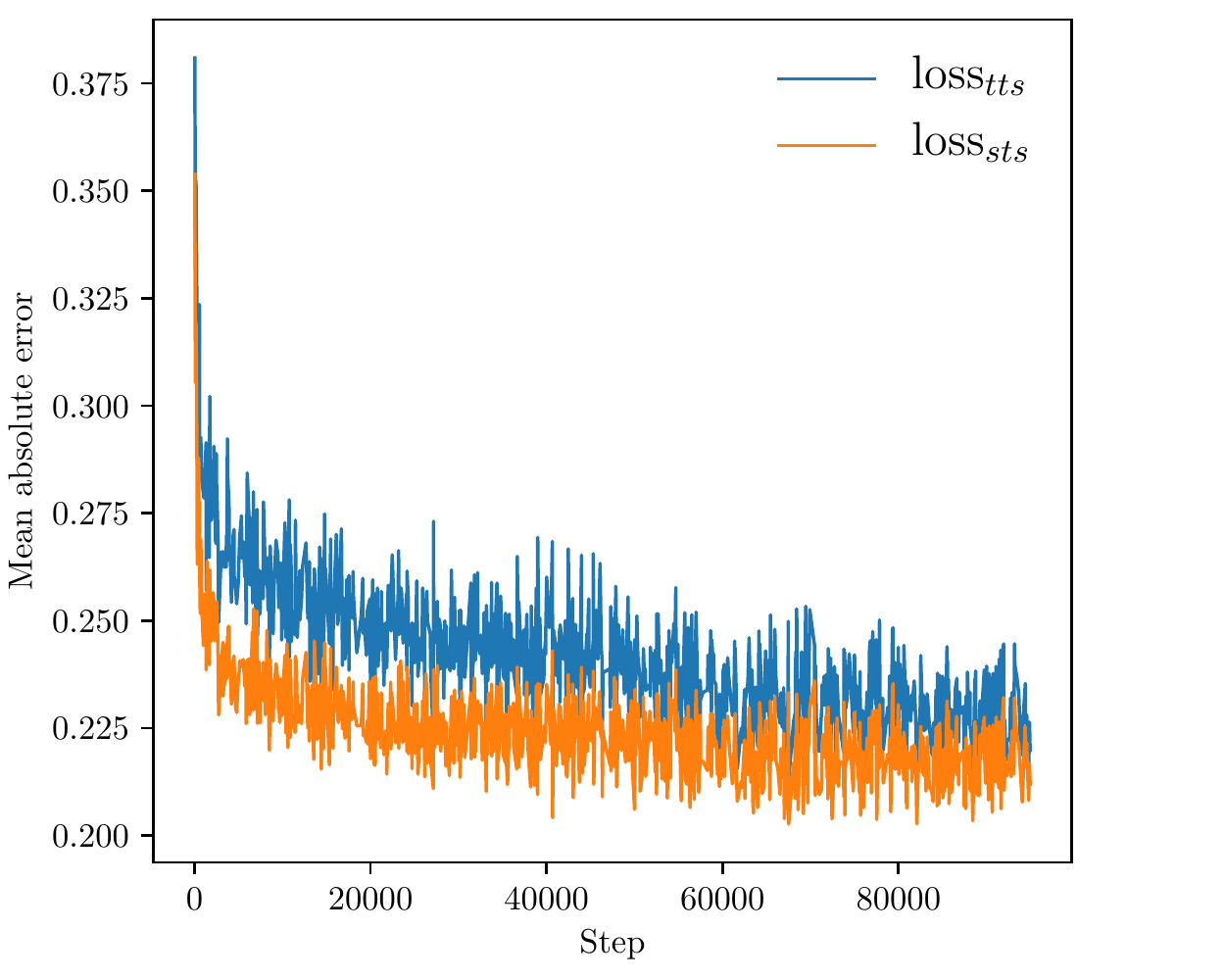}}
  \hfill
  \subfloat[Latent space tying\label{fig:loss-kld}]{\includegraphics[width=0.32\linewidth]{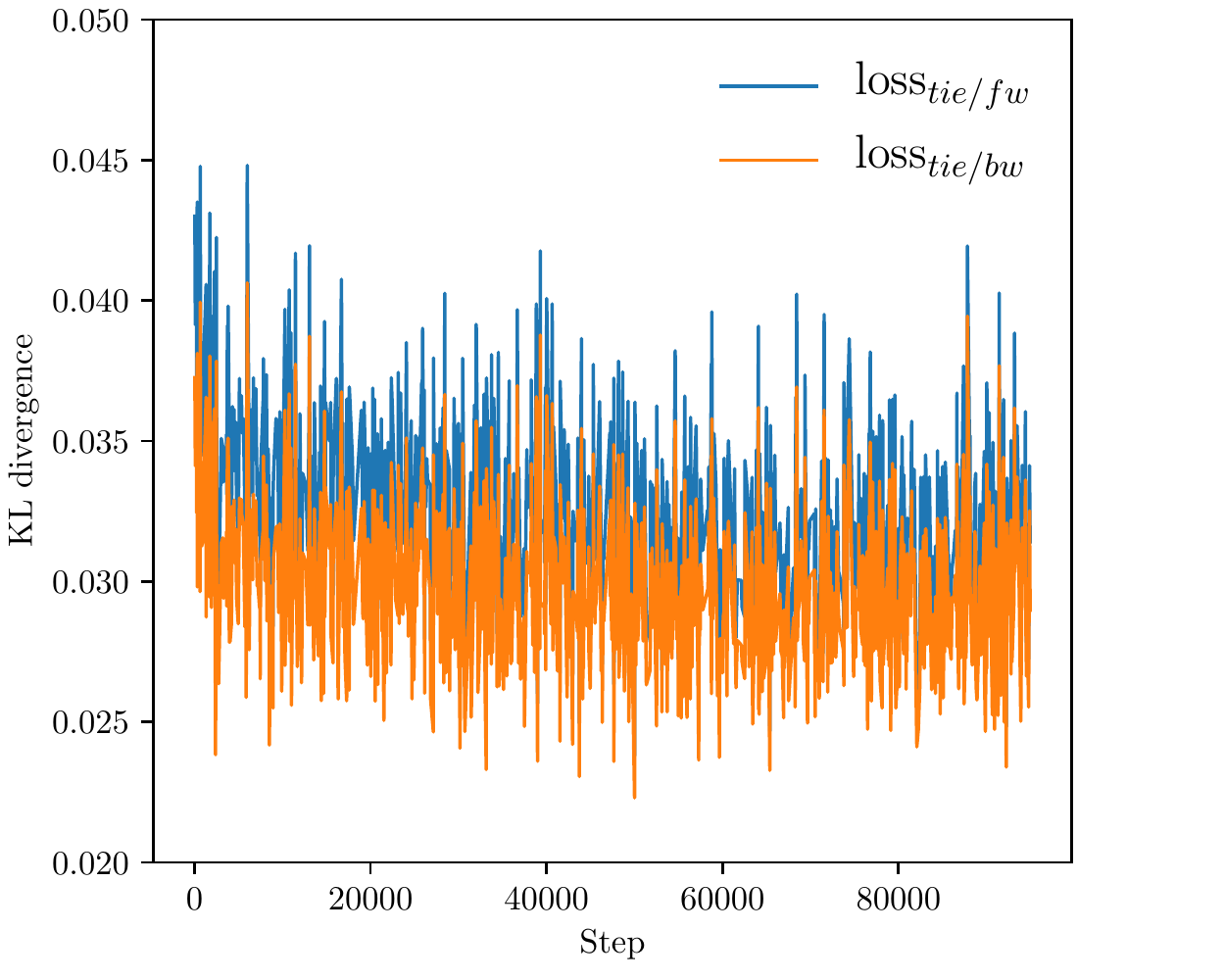}}
  \hfill
  \subfloat[Phoneme classification goals\label{fig:loss-ce}]{\includegraphics[width=0.32\linewidth]{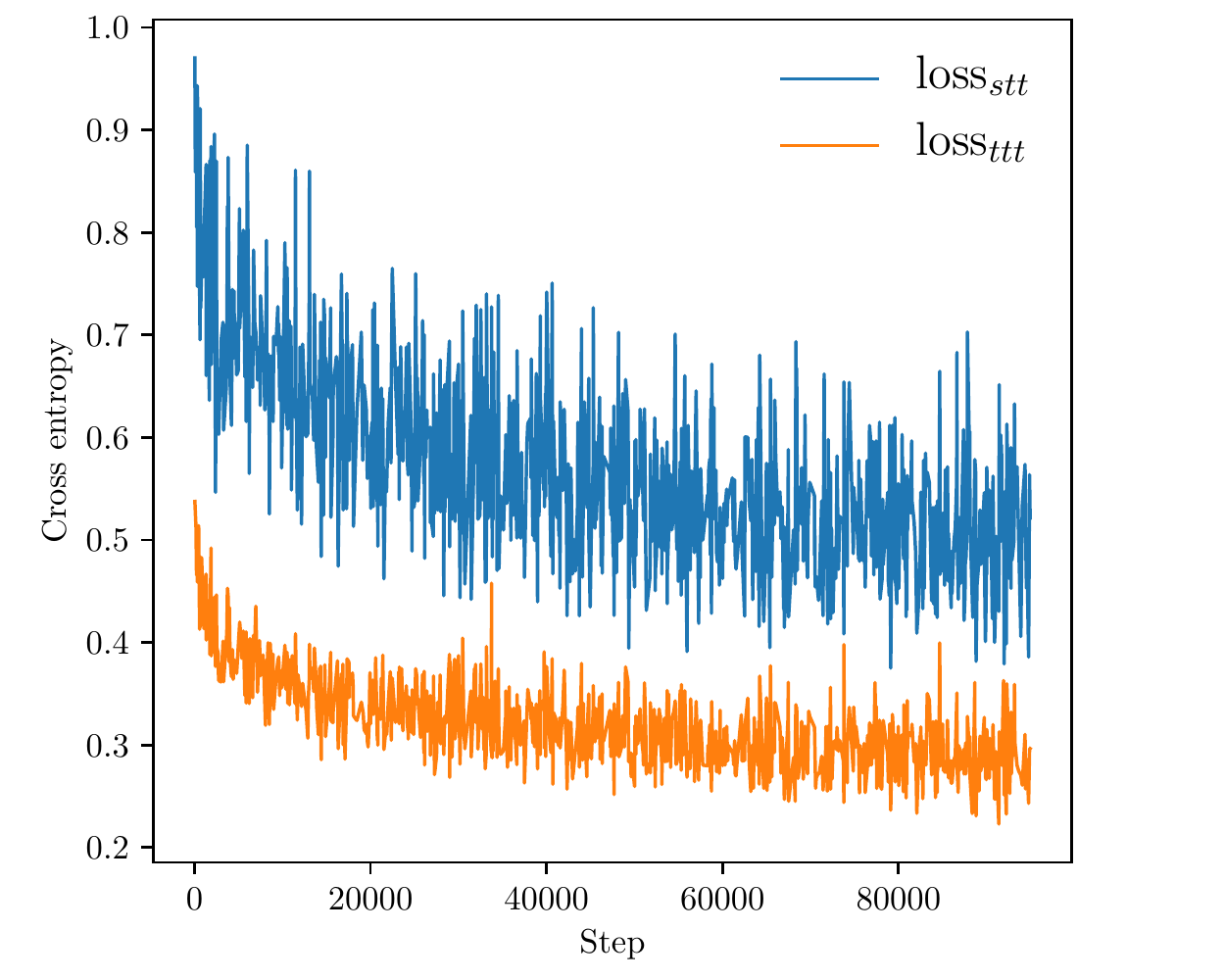}}

\caption{Training curves of different losses available in the training stage of the text-speech multimodal system. The data points are from the training of the model used in scenario A.}
\label{fig:training-losses}
\vspace{-3mm}
\end{figure*}

For the standard case with native target speakers, the subjective results show high MOS scores for most systems as shown in Fig. \ref{fig:exp2-subjective-native}. The new results here are comparisons between supervised and unsupervised approaches. Comparing the XV and FT systems, which represent unsupervised and supervised TTS baselines, we see that the fine-tuned one was significantly better than the speaker embedding one as it benefited from all ten minutes of data. Similar to scenario A, XV system has better WER than FT for many targets.
Among our systems, the difference between the supervised and unsupervised strategies was marginal, but they were all better than the supervised baseline FT. One hypothesis is that our approaches are less sensitive to overfitting thanks to the multi-speaker corpus, speaker factorization and denoising training while FT has a higher possibility of overfitting when using ten minutes of speech  \cite{inoue2020semi,huang2019voice}. These observations are also supported by AB-preference tests (See the bottom part of Fig. \ref{fig:exp2-subjective-native}).

For the challenging case with non-native target speakers, the subjective results revealed more interesting tendencies (Fig. \ref{fig:exp2-subjective-nonnative}). This scenario not only showed the robustness of the voice cloning methods but also the listeners' behaviors. First, we can see that our systems had higher similarity scores than the TTS baselines, FT and XV. The differences between our supervised and unsupervised strategies was more profound in the non-native cases. TTS$_s$ seemed to have higher similarity than TTS$_u$. 
Next, we see that the natural speech of the non-native speakers (NAT) had lower quality scores than their native counterpart. This would be because our native listeners perceived the ``quality'' of speech with strong non-native accents as low, which made the quality and similarity results in this case no longer a positive correlation. The average per-listener results for non-native NAT are plotted in Fig. \ref{fig:exp2-subjective-nat}. A negative correlation was found even for the subjective results of the TTS baselines, FT and XV, indicating that higher-quality speech corresponded to less accented speech and hence lower speaker similarity to non-native target speakers. This highlights the pros and cons of these adaptation methods.
Interestingly, WER of TTS$_s$ was worse than that of TTS$_u$ while the natural speech (NAT) had even worse score in the non-native case. This can be interpreted as that TTS$_s$ produces pronunciation which is more similar to the natural speech than TTS$_u$, which means TTS$_s$ is better at capturing non-standard speaker characteristics.

In summary, the proposed system had higher speaker similarity than the baseline systems. Our TTS system, in particular, benefited from the supervised strategy although the improvement was relatively small. Regarding the TTS$_u$ and the other two VC systems that had slightly better quality than the natural speech, we suspect that this is due to the reduced/lack of accents of their generated speech. This hints at potential uses for other accent-related tasks \cite{aryal2014can}.

\subsubsection{Scenario conclusion}
The subjective results have shown that the fine-tuning approach is better at capturing unique speaker characteristics than the speaker embedding approach when data are sufficient. Our systems, in particular, achieved high performance for native speakers as well as non-native speakers.
Moreover our cloning strategy can be adjusted to take advantage of the transcriptions if they are available. 
In the meantime, the experiment also points out the limitations of the subjective evaluation. While the current quality and similarity questions work well for native speakers, listeners' judgements were biased when they needed to evaluate the voices of non-native speakers.

\section{Analysis and Discussion}

\subsection{Training robust and consistent linguistic latent spaces}
The linguistic latent spaces obtained in the initial training stage have critical effect on the performance of the proposed system, as the rest of the voice cloning procedure functions on the assumption that LLE is a speaker-disentangled linguistic feature. Therefore, the training of the text-speech multimodal system must be carefully designed to guarantee that objective.
If we only consider the text encoder and the speech decoder, then the proposed system is just a multi-speaker TTS model which lacks the ability to adapt with untranscribed speech. By adding a speaker-independent speech encoder, we provide a backdoor for unsupervised speaker adaptation, which is the topic explored in our previous publications \cite{luong2018multimodal,luong2019unified}.
If we only consider the speech encoder and speech decoder, then it is not much different from a VAE-based multi-speaker non-parallel VC system \cite{hsu2016voice}.
However, to avoid the weakness of self-supervised models, which is the dependence on regularization to shape the latent space indirectly, we jointly trained the STS stack with the text encoder and transcribed speech in a supervised fashion.
This ensures that the latent spaces will contain linguistic information which in turn guarantees a high performance for VC \cite{luong2019bootstrapping}.

\begin{table}[tb]
    \caption{Description and word error rates of evaluated setups.}
    \scalebox{0.95}{
    \begin{tabular}{llrrrr}
        \hline \hline
        Setup & Denote & \multicolumn{2}{c}{TTS$_u$} & \multicolumn{2}{c}{VCM$_u$} \\
         & & p294 & p345 & p294 & p345 \\\hline
        NAUTILUS & N & 8.37 & 9.74 & 22.22 & 24.05  \\ \hline
        N - welding & A & 8.98 & 9.44 & 23.90 & 24.35 \\ 
        N - $\textrm{loss}_{cycle}$ in adaptation & B & 8.98 & 11.11 & 22.68 & 22.98 \\ 
        N - text decoder in training & C & 12.33 & 12.02 & 25.57 & 25.88 \\ 
        N - all above (A, B \& C) & D & 10.35 & 13.85 & 27.55 & 29.38 \\ \hline
    \end{tabular}}
    \label{tab:ana-wer}
\vspace{-3mm}
\end{table}

\begin{figure*}[t]
\centering
    \subfloat[Native speaker as target - p294\label{fig:uttlle_p294}]{\includegraphics[width=0.5\linewidth]{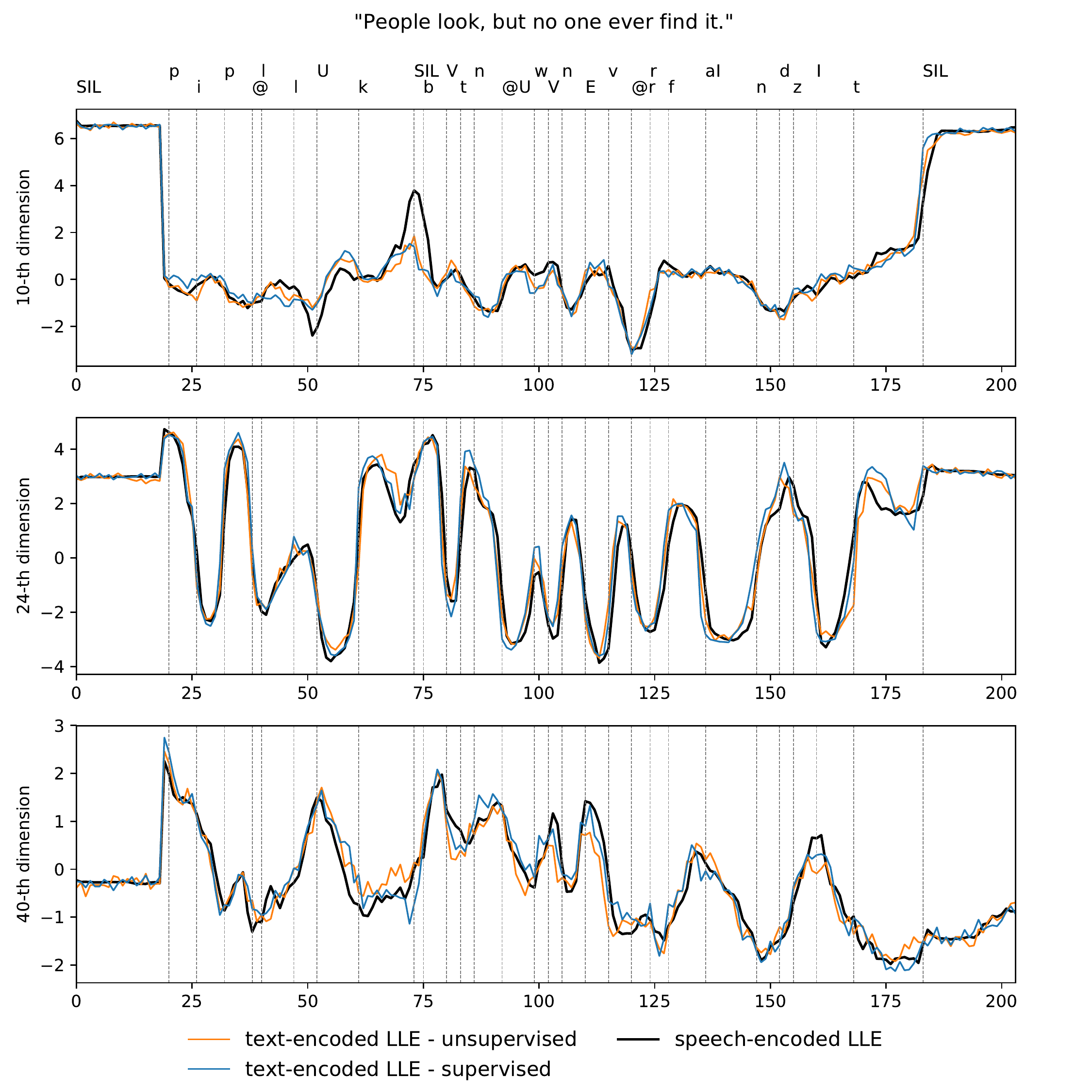}}
    \hfill
    \subfloat[Non-native speaker as target - MF6\label{fig:uttlle_MF6}]{\includegraphics[width=0.5\linewidth]{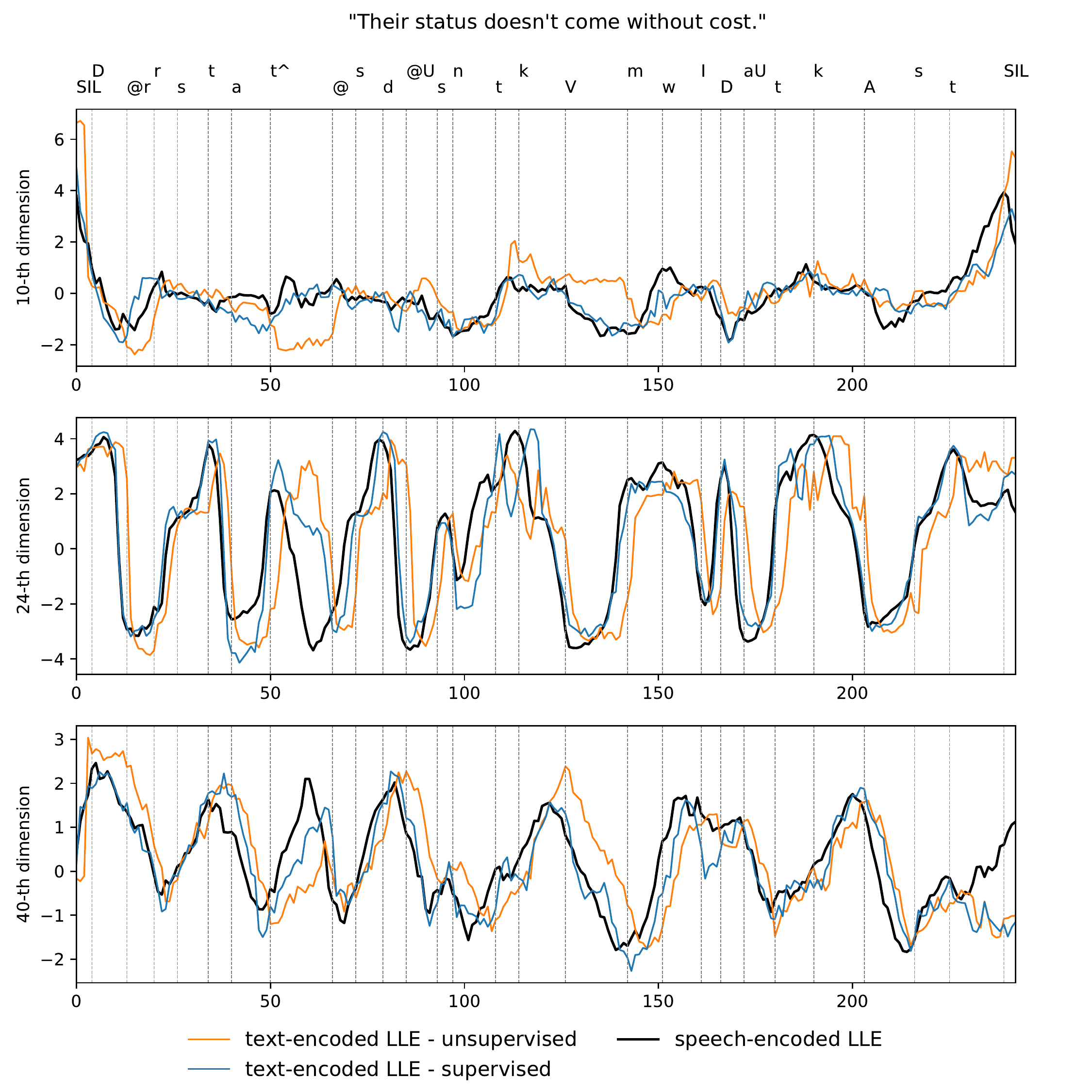}}
    \label{fig:explore-lle}
    \caption{Examples of 64-dimensional LLE sequences generated by the text and speech encoders of models adapted using either the supervised or unsupervised cloning strategy. An utterance of the target speaker was used to generate the speech-encoded LLE with the speech encoder, while text (phoneme) and alignment information extracted from the same utterance were used to generate the text-encoded LLE with either the supervised or unsupervised text encoder.}
\vspace{-3mm}
\label{fig:uttlle}
\end{figure*}

By jointly training the text and speech encoder, we help the speech encoder to learn a speaker-disentangled representation, as it is forced to approximate the text encoder, which is speaker-independent by nature.
Figure\ \ref{fig:loss-mae} shows the training curves of the TTS and STS goals, both of which descend over time and gradually converge to each other. In practice, we have to de-emphasize $\textrm{loss}_{sts}$ with a weighting parameter and observe the progress of the training curves, as there is a risk that the training will focus on optimizing $\textrm{loss}_{sts}$ and abandon $\textrm{loss}_{tts}$ completely, as there is always an easy and uninteresting solution to the autoencoder task.
In summary, a robust and speaker-disentangled latent linguistic representation is guaranteed by strategic placement of speaker components, joint training of the TTS and STS stacks, and use of a large-scale transcribed multi-speaker corpus.
Furthermore, the tied-layer loss is used in conjunction with the join-goal losses to encourage a consistent representation between text-encoded and speech-encoded latent spaces.
Figure\ \ref{fig:loss-kld} shows the forward and backward KL divergence learning curves, which reveals that a small gap still exists between the two.
Finally, the text decoder was used to force the LLE to focus more on phonetic information by adding $\textrm{loss}_{stt}$ to the optimizing loss.
Interestingly, even though $\textrm{loss}_{ttt}$ is not optimized, it is still better than $\textrm{loss}_{stt}$, as can be seen in Fig.\ \ref{fig:loss-ce}.

\subsection{Effects of auxiliary techniques on word error rates}

Beside the new architecture (Section \ref{sec:condition}) and the large-scale training corpus, which are the main contributors to the improved performance compared with our previous work \cite{luong2019unified,luong2019bootstrapping}, the NAUTILUS system also incorporates many new auxiliary/optionally techniques.
In this section, we investigate their effect on the word error rate of generated speech samples.
Specifically, several slightly different setups of the proposed system were evaluated in the unsupervised speaker adaptation scenario. The experimental environment of scenario B was reused, but we only evaluated the two native English speakers, as their results are easier to interpret.
Table \ref{tab:ana-wer} lists the WER of the generated utterances produced by different setups. Setup N is the unsupervised voice cloning process described in Section \ref{sssec:adaptation-process}, A does not include the welding step, B removes $\textrm{loss}_{cycle}$ from the adaptation loss, C is not trained with the text decoder, while D removes all three elements from the procedure. In the other words, D is the most similar to the setup used in our previous publication \cite{luong2019unified,luong2019bootstrapping}.
Interestingly, setup A and B have WER not much different from N, while setup C and D are significantly worse than the others even though they were all trained and adapted on the same data. These results suggest that the text decoder plays a significant role in improving the pronunciations of the generated utterances.
By comparing setup C and D, we can see that the welding step and the linguistic cycle consistent also have positive impact on WER but their effects are smaller and more situational.

Having these complementary techniques at disposal can be useful for squeezing out the last bit of performance in production.
If we have reliable automatic metrics, the cloning strategy can be personalized to accommodate a specific target speaker or a specific application scenario by searching for the optimal setup and hyperparameters for the particular situation, which is a topic that we will explore more in our future works. The speech samples of these setup can also be found in the accommodated web page.

\subsection{LLE of the supervised and unsupervised models}

As mentioned in earlier sections, the architecture of the NAUTILUS system used in this paper is not an E2E system, which is inconvenient for practical applications but it allows us to have more control over the duration of the generated utterances.
In this section, we look into the linguistic latent spaces of the adapted models to understand the behaviors of the supervised and unsupervised cloning strategies.
Figure\ \ref{fig:uttlle} shows selected dimensions of the 64-dimensional LLE sequences generated by either the text or speech encoder of models adapted to either p294 or MF6. For each target speaker, we used speaker-independent speech encoder and an utterance not included in their adaptation data to generate a speech-encoded LLE sequence and then used either the supervised or unsupervised text encoder and the phoneme (text) and duration information of the same utterance to generate text-encoded LLE sequences.
This arrangement guarantees the LLE sequences generated from the encoders are aligned, which helps to highlight differences between the supervised and unsupervised text encoders.

Even though we referred to the outputs of both the text and speech decoder as LLE, they actually represent slightly different concepts. The speech-encoded LLE represents the sound spoken in an utterance input, while the text-encoded LLE represents the sound that we want to generate from a symbolic phoneme input.
Figure\ \ref{fig:uttlle_p294} shows all three LLE sequences are well-aligned to each other in the case of p294.
This suggests that the unsupervised speaker-independent text encoder was able to correctly map the symbolic phoneme to the actual spoken sound when the target is a native English speaker, which left little room for the supervised strategy to improve upon. It is expected as the text and speech encoders were initially trained on transcribed speech of a large-scale native speaker corpus.
In contrast, Fig.\ \ref{fig:uttlle_MF6} shows clear misalignments between the LLE sequences; the text-encoded LLE sequence of the supervised model seems to align to the speech-encoded LLE sequence better than its unsupervised counterpart.
From this figure, we can see that the supervised strategy adjusted the text encoder to map the symbolic phoneme to the actual (wrong) sound spoken by MF6, which helps to improve the speaker similarity but degrades the quality (or pronunciation) of generated utterances. The latent spaces of the models adapted to p345 and MM6, while not presented in this paper, also show similar patterns.

\section{Conclusion}
\label{sec:conclusion}
In this paper, we showed that our voice cloning system, ``NAUTILUS'', can achieve state-of-the-art performance.
More importantly, it can act as a text-to-speech or voice conversion system with high consistency in terms of speaker characteristics when switching between the two.
With the versatile cloning strategy, which can be adjusted to specific data situation of a target speaker, it is potentially useful for many other interesting tasks like accent reduction \cite{aryal2014can} or cross-lingual voice cloning \cite{abe1991statistical,zhou2019cross}.
For future work, we will focus on evaluating our systems by using different architectures for text-speech systems \cite{li2019neural,huang2019voice} or neural vocoders \cite{mehri2016samplernn,wang2019neural} to solve specific voice cloning scenarios \cite{zhang2019learning,luong2019bootstrapping}.
Finally given the multimodal structure, extending our system to other speech generation tasks (e.g., video-to-speech \cite{cornu2015reconstructing}) would be a natural direction toward a unified voice cloning framework.

\section*{Acknowledgments}

This work was partially supported by a JST CREST Grant (JPMJCR18A6, VoicePersonae project), Japan, and MEXT KAKENHI Grants (16H06302, 17H04687, 18H04120, 18H04112, 18KT0051), Japan. 

\ifCLASSOPTIONcaptionsoff
  \newpage
\fi



%

\bibliographystyle{IEEEtran}
\bibliography{main}

\newpage
\begin{IEEEbiography}{Hieu-Thi Luong}
received a Ph.D. degree from the Graduate University for Advanced Studies (SOKENDAI), Japan, in 2020 for a thesis focuses on unifying the methodology of cloning voices using text-to-speech and voice conversion systems.
He received B.E and M.E degrees in computer science from Vietnam National University, Ho Chi Minh City, University of Science, Vietnam in 2014 and 2016 respectively while working on speech technology systems for Vietnamese language. In 2017, he was awarded a Japanese Government (Monbukagakusho: MEXT) Scholarship to pursue a PhD degree in statistical speech synthesis and machine learning at the National Institute of Informatics, Tokyo, Japan.
\end{IEEEbiography}

\begin{IEEEbiography}{Junichi Yamagishi}
received a Ph.D. degree from The Tokyo Institute of Technology in 2006 for a thesis that pioneered speaker-adaptive speech synthesis. He is currently a Professor with the National Institute of Informatics, Tokyo, Japan and also a Senior Research Fellow with the Centre for Speech Technology Research, University of Edinburgh, Edinburgh, U.K. Since 2006, he has authored and co-authored more than 250 refereed papers in international journals and conferences.

He was the recipient of the Tejima Prize as the best Ph.D. thesis of Tokyo Institute of Technology in 2007. He was awarded the Itakura Prize from the Acoustic Society of Japan in 2010, the Kiyasu Special Industrial Achievement Award from the Information Processing Society of Japan in 2013, the Young Scientists' Prize from the Minister of Education, Science and Technology in 2014, the JSPS Prize from the Japan Society for the Promotion of Science in 2016, and the Docomo Mobile Science Award from the Mobile Communication Fund in 2018.  

He was one of the organizers for the special sessions on ``Spoofing and Countermeasures for Automatic Speaker Verification" at Interspeech 2013, ``ASVspoof evaluation" at Interspeech 2015, ``Voice Conversion Challenge 2016" at Interspeech 2016, ``2nd ASVspoof evaluation" at Interspeech 2017, and ``Voice Conversion Challenge 2018” at Speaker Odyssey 2018. He is currently an organizing committee member for ASVspoof 2019, an organizing committee member for the 10th ISCA Speech Synthesis Workshop 2019, a technical program committee member for IEEE ASRU 2019, and an award committee member for ISCA Speaker Odyssey 2020.

He was a member of the Speech and Language Technical Committee and a Lead Guest Editor for a special issue of the IEEE Journal of Selected Topics in Signal Processing on spoofing and countermeasures for automatic speaker verification. He is currently a Senior Area Editor of the IEEE/ACM Transactions on Audio, Speech, and Language Processing and a chairperson of ISCA SynSIG.
\end{IEEEbiography}

\end{document}